\documentclass[aps,12pt,superscriptaddress,tightenlines,preprint,floatfix]{revtex4-1}
\usepackage[dvips]{graphicx}
\usepackage{amsmath,amssymb,amsthm,bm,multirow,longtable}
\newcommand{\nn}{\nonumber}
\renewcommand{\epsilon}{\varepsilon}
\renewcommand{\phi}{\varphi}
\newcommand{\GL}{\text{GL}}
\newcommand{\C}{\mathbb{C}}
\newcommand{\N}{\mathbb{N}}
\newcommand{\R}{\mathbb{R}}
\DeclareMathOperator{\im}{im}
\DeclareMathOperator{\id}{id}

\DeclareMathOperator{\linearspan}{span}
\providecommand{\abs}[1]{\lvert#1\rvert}
\theoremstyle{plain}

\newtheorem{theorem}{Theorem}
\theoremstyle{definition}
\allowdisplaybreaks

\begin{document}
\title{An algebraic classification of entangled states}
\author{Roman V. Buniy}
\email{roman.buniy@gmail.com}
\affiliation{School of Earth and Space Exploration, Arizona State University, Tempe, AZ 85287}
\affiliation{Chapman University, Schmid College of Science, Orange, CA 92866}
\altaffiliation{Permanent address}
\author{Thomas W. Kephart}
\email{tom.kephart@gmail.com}
\affiliation{Department of Physics and Astronomy, Vanderbilt University, Nashville, TN 37235}
\date{\today}
\begin{abstract}
  We provide a classification of entangled states that uses new discrete entanglement invariants.
  The invariants are defined by algebraic properties of linear maps associated with the states.
  We prove a theorem on a correspondence between the invariants and sets of equivalent classes of entangled states.
  The new method works for an arbitrary finite number of finite-dimensional state subspaces.
  As an application of the method, we considered a large selection of cases of three subspaces of various dimensions.
  We also obtain an entanglement classification of four qubits, where we find 27 fundamental sets of classes.
\end{abstract}
\maketitle

\section{Introduction}

The phenomenon of entanglement is one of the most fundamental and counterintuitive features of quantum mechanics.
Its fundamental role was emphasized by the formulation of the EPR paradox \cite{EPR}, despite the original purpose of the latter to question physical reality of the wave function.
The counterintuitive nature of entanglement is a hallmark of quantum mechanics, and its properties reveal deep distinctions between quantum and classical objects.

From the mathematical point of view entanglement is a consequence of the superposition principle and the tensor product postulate in quantum mechanics.
Specifically, the principle and postulate imply that a state vector of a system consisting of several subsystems is a linear combination of tensor products of state vectors of the subsystems.
Mathematical and physical properties of states interrelate: a state is disentangled if it can be transformed into a factorizable state; any other state is entangled. 
Equivalently, a state is disentangled if and only if each subsystem is in a definite state.

Despite the simplicity of the above qualitative features of entanglement the complete list of its quantitative characteristics is unknown.
For example, it might appear that the smallest number of linearly independent factorizable terms representing a state is an appropriate characteristic of its entanglement.
This is true for two subsystems, in which case this single quantity completely classifies all entangled states.
For more than two subsystems, however, this quantity does not characterize entanglement since it depends on a choice of bases.
To choose appropriate entanglement characteristics for the general case we need to study invariant properties of states of composite systems; these are the key properties shaping the following discussion.

For states of composite systems entanglement quantifies ways in which states of subsystems contribute to linear combinations of tensor products.
The larger the numbers of contributing states of subsystems, the greater the variety of arrangements of terms in linear combinations.
However, some of the arrangements should be considered as dependent since they are related by transformations of bases.
Such related states form equivalence classes, finding the complete set of which is the goal of entanglement classification.  

To classify entangled states one usually employs entanglement invariants, which are certain invariant quantities associated with the states.
The nature of the problem requires that the invariants do not change under all transformations that can be reduced to changes of bases.
Consequently, the invariants take the same values for all states within each equivalence class, and the standard method of finding them uses the classical theory of invariants \cite{Olver}.
Variants of this method are used in most known cases of partial or complete entanglement classification; see, for example, \cite{Gelfand,Dur,Verstraete,Klyachko,Miyake1,Luque1,Miyake2,Briand,Luque2,Toumazet,Wallach,Lamata,Cao,Li,Akhtarshenas,Chterental,Borsten:2008wd,Borsten:2010db}.

Developing the ideas outlined above, we have introduced in \cite{Buniy:2010yh} a new entanglement classification method based on algebraic properties of tensor products of linear maps.
In this paper we generalize and expand both the method and its applications.
We first introduce various equivalence relations and corresponding equivalence classes on linear spaces of states.
We then show how these classes lead to various linear subspaces and their invariants, which are the central objects in our method of algebraic classification of entangled states.
During the development of our method we turn repeatedly to the example of three qubits to illustrate the procedure.
Finally, we proceed with numerous more complicated but physical relevant examples demonstrating the use of the method in classifying the entanglement of many systems unsolved until now.

\section{Preliminaries}

\subsection*{Tensor product space}

We begin by introducing the main components of our construction.
Let $S$ be a quantum system that consists of $n$ subsystems $\{S_i\}_{i\in I}$, where $I=\{1,\dotsc,n\}$.
For each $i\in I$, let a finite-dimensional vector space $V_i$ over a field $F$ be the state space of $S_i$.
Extension to infinite-dimensional spaces is nontrivial and is not considered here.
We choose $F=\R$ for the simplicity of presentation; the case $F=\C$ needs only minimal modifications.

Our first task is to define the space $V$, the state space of $S$.
The tensor product postulate in quantum mechanics says that $V$ is a subspace of the tensor product space $\otimes_{i\in I}V_i$.
A specific choice of subspace $V$ depends on the nature of $S$.
For identical subsystems, for example, the permutation symmetry acting on the subsystems determines $V$.
In particular, for bosonic or fermionic subsystems $V$ is, respectively, the symmetric or antisymmetric part of the product $\otimes_{i\in I}V_i$.
Also, if there is an equivalence relation among elements of $V$ (as, for example, for linearly dependent vectors in quantum mechanics), then $V$ is the appropriate quotient set.
Modifications due to these and similar properties can be easily included into the following development, which assumes the simplest case where $V=\otimes_{i\in I}V_i$.

\subsection*{Transformation group}

We aim to study properties of the system $S$ related to its composition in terms of the subsystems $\{S_i\}_{i\in I}$; these are equivalent to properties of $V$ related to its composition in terms of $\{V_i\}_{i\in I}$.
The latter manifest themselves in their transformations under an appropriate group.
Note that the tensor structure of $V$ implies that the transformation group relevant for studying properties of $V$ is not the general linear group of $V$, $\GL(V)$, but rather its subgroup $\times_{i\in I}\GL(V_i)$.
Accordingly, for each $i\in I$ we choose a subgroup $G_i$ of $\GL(V_i)$ and define the corresponding subgroup $G=\times_{i\in I}G_i$ of $\GL(V)$.
As a result, the group $G$ is the transformation group for $V$, and it determines properties of $V$ related to its composition in terms of $\{V_i\}_{i\in I}$.
Particular cases (where only certain subsets of $V$ and subgroups of $G$ matter) are of interest as well and can be treated similarly to the general case of $V=\otimes_{i\in I}V_i$ and $G=\times_{i\in I}\GL(V_i)$ considered here.

\subsection*{Equivalence classes}

The group $G$ induces the equivalence relation $\sim_{V}$ on $V$, which is given by $v'\sim_{V}v$ for each $v,v'\in V$ if and only if there exists $g\in G$ such that $v'=gv$.
The equivalence relation defines the equivalence class of $v$ under $\sim_{V}$,
\begin{align*}
  C(v)=\{v'\in V \colon v'\sim_{V}v\}.
\end{align*}
Since all elements of the class $C(v)$ are equivalent, any one of its elements determines the whole class.
It is thus convenient to replace $C(v)$ with its arbitrary single element $\tilde{v}\in C(v)$, which we call a representative element of the class.
(For each specific class $C(v)$ the choice of $\tilde{v}$ based on various symmetry considerations generally leads to simplifications.)
Repeating this procedure for each $v\in V$, we partition $V$ into the set of equivalence classes
\begin{align*}
  C=\cup_{v\in V}\{C(v)\}
\end{align*}
such that each vector in $V$ belongs to one and only one class.
Finally, replacing each class in $C$ by its representative element, we arrive at the set
\begin{align*}
  \tilde{V}=\{\tilde{v}\in C(v)\colon C(v)\in C\},
\end{align*}
which can also be written as the quotient set $\tilde{V}=V/\sim_{V}$.

\subsection*{Properties of entangled states}

Understanding the structure of the quotient set $\tilde{V}$ is our ultimate goal.
We begin with a general property of $\tilde{V}$, its partition into three characteristic subsets of vectors: (1) the zero vector, (2) decomposable vectors, (3) nondecomposable vectors.
By definition, a decomposable vector $v\in V$ is a vector that can be written in the factorizable form $v=\otimes_{i\in I}v_i$, where $v_i\in V_i$ is a nonzero vector for each $i\in I$.
A nondecomposable vector is a vector which is neither zero nor decomposable.
We will derive the general form of a nondecomposable vector after we establish its invariant characteristics.

The above partition is physically significant because it is in a one-to-one correspondence with the partition of quantum states into three types: (1) the vacuum state, (2) disentangled states, (3) entangled states.
The zero vector (the vacuum state) and decomposable vectors (disentangled states) are the simplest elements of $V$; although they comprise only a small part of $V$, they span all of it.
By contrast, nondecomposable vectors (entangled states) are more complex and difficult to categorize.
The difficulty is combinatorial because decomposable vectors from $V$ that enter the linear combination representing a nondecomposable vector differ by ways in which linearly independent vectors from $\{V_i\}_{i\in I}$ enter the expression.
Finding all such possibilities of nonequivalent combinations (which is the same as finding the quotient set $\tilde{V}$) is the problem of entanglement classification.

Another general property of $\tilde{V}$ concerns the number of its elements.
Although the set $\tilde{V}$ is not a vector space, we use the notation $\dim{\tilde{V}}$ for the number of unconstrained elements of $F$ that a general element of $\tilde{V}$ depends on.
Using a similar notation for $\dim{G}$, we find
\begin{align*}
  \dim{\tilde{V}}\ge\dim{V}-\dim{G}.
\end{align*}
The inequality sign appears here because, in general, the system of linear equations for $g\in G$ that follows from the equivalence condition $v'=gv$ is not linearly independent.
We have two distinct cases here: (1) if $\dim{V}-\dim{G}\le 0$, the above inequality does not tell us if there are any unconstrained elements of $F$ that a general element of $\tilde{V}$ depends on; (2) if $\dim{V}-\dim{G}>0$, there are at least $\dim{V}-\dim{G}$ such elements of $F$.
Consequently, $\tilde{V}$ is an infinite set in the second case.
Asymptotically for large $n$, $\dim{V}$ is exponential in $n$ and $\dim{G}$ is at most quadratic in $n$.
It follows that $n$ does not need to be very large for the set $\tilde{V}$ to be infinite; in other words, $\tilde{V}$ is typically infinite.

\subsection*{Example of three qubits}

To illustrate the concepts introduced above, we consider a particular example of three qubits, in which case there are three $2$-dimensional spaces $V_1,V_2,V_3$ and their tensor product $V=V_1\otimes V_2\otimes V_3$.
We choose arbitrary bases $\{e_{i,j}\}_{1\le j\le 2}$ for each $V_i$ and expand an arbitrary element $v\in V$ in terms of its coordinates $\{v_{j_1,j_2,j_3}\}$, 
\begin{align*}
  v=\sum_{j_1=1}^2\sum_{j_2=1}^2\sum_{j_3=1}^2 v_{j_1,j_2,j_3} e_{1,j_1}\otimes e_{2,j_2}\otimes e_{3,j_3}.
\end{align*}
The transformation group $G$ acts on $v$ according to $v\mapsto g_1 g_2 g_3 v$, where $g_i\in\GL(V_i)$ and the coordinates of $g_1 g_2 g_3 v$ are
\begin{align*}
  (g_1 g_2 g_3 v)_{j_1,j_2,j_3}=\sum_{k_1=1}^2\sum_{k_2=1}^2\sum_{k_3=1}^2 v_{k_1,k_2,k_3}(g_1)_{k_1,j_1}(g_2)_{k_2,j_2}(g_3)_{k_3,j_3}.
\end{align*}

The simplest example of a decomposable vector in $V$ (a disentangled state in $S$) is $v=e_{1,1}\otimes e_{2,1}\otimes e_{3,1}$.
This is a state of the system $S$ in which each of its subsystems $S_i$ is in a definite state $e_{i,1}$.

There are three types of nondecomposable vectors in $V$ (entangled states in $S$). 
For the first type, vectors are nondecomposable for the tensor product of two spaces but decomposable for the tensor product of the three spaces.
Choosing $V_1$ and $V_2$ as two such spaces, we have the state $v=(e_{1,1}\otimes e_{2,1}+e_{1,2}\otimes e_{2,2})\otimes e_{3,1}$ in which the subsystems $S_1$ and $S_2$ are not in definite states, while the subsystem $S_3$ is in a definite state. 
The other two states of this type are obtained by permutation of the subsystems.
For the second and third type (which in the literature are called respectively the W and the GHZ classes), states are nondecomposable for the tensor product of the three spaces.
The standard forms of their representative states are $v=e_{1,1}\otimes e_{2,1}\otimes e_{3,2}+e_{1,1}\otimes e_{2,2}\otimes e_{3,1}+e_{1,2}\otimes e_{2,1}\otimes e_{3,1}$ and $v=e_{1,1}\otimes e_{2,1}\otimes e_{3,1}+e_{1,2}\otimes e_{2,2}\otimes e_{3,2}$, respectively.
For these states, no subsystem is in a definite state. 

Applying all elements of the group $G$ to a representative vector $v$ in any of the above three types of elements of $V$, we obtain the equivalence class $C(v)$, which leads to $7$ equivalence classes (counting permutations and including the zero vector which is in its own equivalence class).
This is a well-known result (which we also proved by using our method in \cite{Buniy:2010yh}) that these $7$ classes constitute the complete entanglement classification of three qubits.

\subsection*{Invariants}

The problem of finding $\tilde{V}$ can be solved by direct or indirect methods.
In a direct method, one uses the definition of $\tilde{V}$ to derive the general form of representative elements of equivalence classes.
Although there are no restrictions to such methods in theory, they are usually inefficient in practice because of the need to solve complicated equations.
By contrast, in an indirect method, one seeks quantities characterizing elements of $V$ which are invariant under $G$.
Equivalence classes are obtained by finding allowed values of these invariants.
Indirect methods are usually efficient if all invariants are known.

Continuing with indirect methods, let $a(v)\in F$ be an invariant of $v$ induced by the group $G$.
This is a quantity that satisfies $a(v')=a(v)$ for each $v\in V$, $v'\in C(v)$, which implies that invariants depend only on classes.
Let $A(v)$ be a complete set of algebraically independent invariants of $v$, so that $v'\sim_{V}v$ if and only if $A(v')=A(v)$, for each $v,v'\in V$.
The standard method of finding $A(v)$ is to use the classical theory of invariants and covariants; for a modern introduction, see, for example, \cite{Olver}.
Almost all known cases of partial or complete entanglement classifications use this method to a certain extent; see, for example, \cite{Gelfand,Dur,Verstraete,Klyachko,Miyake1,Luque1,Miyake2,Briand,Luque2,Toumazet,Wallach,Lamata,Cao,Li,Akhtarshenas,Chterental,Borsten:2008wd,Borsten:2010db}.
The rapid increase of $\abs{A(v)}$ with $n$ is the main reason why only the simplest cases of entanglement classification have been fully carried out.

Let us now consider a typical case of infinite $\tilde{V}$.
We find that the set of all possible values of the invariants, $\cup_{v\in V}\{A(v)\}$, is infinite.
The resulting information about $\tilde{V}$ in terms of its elements and invariants is both overwhelming in its detail and impractical in its use.
As a key part of our method, we reduce the amount of information by grouping equivalence classes into a finite number of sets.
The grouping is determined by certain equivalence relation between classes in each set, a natural choice for which is defined as follows.

\subsection*{Equivalence of invariants}

We first introduce the rescaling equivalence of invariants.
We note that since linearly dependent vectors in quantum mechanics correspond to the same physical state, we require $fv\in C(v)$ for each $v\in V$, $f\in F$, $f\not=0$.
It follows that algebraic invariants are homogeneous polynomials; consequently, zero is the most important value of each invariant.
This suggests we extend the above rescaling equivalence of states to the rescaling equivalence of invariants.
Specifically, we define the equivalence relation $\sim_F$ on the field $F$ by setting $a'\sim_F a$ for each $a,a'\in F$ if and only if there exists $f\in F$, $f\not=0$ such that $a'=fa$.
(For $F=\R$ or $F=\C$, this simply means that any two nonzero elements are equivalent.)
It is easy to generalize this equivalence to ordered sets over $F$, so that for each pair of such sets $(a'_k)_{k\in K}$ and $(a_k)_{k\in K}$ we define $(a'_k)_{k\in K}\sim_F(a_k)_{k\in K}$ if and only if $a'_k\sim_F a_k$ for each $k\in K$.

Having established equivalence for invariants, we transfer it to vectors.
Namely, we define the equivalence relation $\sim'_{V}$ on the set $V$ by setting $v'\sim'_{V}v$ if and only if $A(v')\sim_F A(v)$, for each $v,v'\in V$.
Since invariants depend only on classes, $v'\sim_{V}v$ implies $v'\sim'_{V}v$.
The relation $\sim_{V}'$ defines the quantities $C'(v)$, $\tilde{v}'$, $C'$, $\tilde{V}'$ in the same manner as the relation $\sim_{V}$ defines the quantities $C(v)$, $\tilde{v}$, $C$, $\tilde{V}$.
Clearly, $C'$ is a partition of $C$.

The sets $C'$ and $\tilde{V}'$ are the main objects of our study.
We call the problem of finding them the restricted entanglement classification problem to emphasize that we seek only sets of classes, not the classes themselves.
One way to solve the problem is to use the set of invariants $A(v)$ from the standard classification method.
This approach requires studying conditions under which elements of $A(v)$ are zero.
If $A(v)$ is known, this method gives the solution; however, we prefer a simpler approach that uses new algebraic invariants $\tilde{N}(v)$ instead of $A(v)$.
The advantage of our approach is that each element of $\tilde{N}(v)$ describes certain algebraic properties of $v$ and takes a value from only a finite set of integers.
The construction of $\tilde{N}(v)$ uses only basic linear algebra~\cite{algebra} and proceeds as follows.

\section{Method}

\subsection*{Outline}

The set of invariants $\tilde{N}(v)$ is uniquely determined by the following conditions.
First, $\tilde{N}(v)$ depends only on the equivalence class $C'(v)$ to which $v$ belongs.
As a result, both $C'(v)$ and $\tilde{N}(v)$ are invariant under the action of the transformation group $G$.
Second, the rescaling property of $A(v)$ implies that $\tilde{N}(v)$ depends only on properties of linear subspaces of $V$; let $L(v)$ be the set of such subspaces.
Third, $L(v)$ depends linearly on $v$.
Fourth, $L(v)$ describes properties of $v$ associated with all partitions of the system $S$ into subsystems build from $\{S_i\}_{i\in I}$.
Such partitions result from all choices of writing $V$ as the tensor product of spaces built from $\{V_i\}_{i\in I}$.

\subsection*{Maps}

The above conditions require that $L(v)$ is defined in terms of linear maps.
We find these as follows.
We first partition the system $S$ into subsystems $T$ and $T'$, so that $S=T\cup T'$.
Let $W$ and $W'$ be the state spaces for $T$ and $T'$, respectively, so that $V=W\otimes W'$.
Our main tool for constructing $L(v)$ is a linear map
\begin{align*}
  f(v)\colon W\to W', \quad f(v)(w)=v\otimes w^*,
\end{align*}
where $w^*\in W^*$ is the dual of $w\in W$.

According to a standard result in linear algebra, all information about a linear map is contained in two fundamental spaces associated with it: its kernel and image,
\begin{align*}
  \ker{f(v)} &= \{w\in W\colon f(v)(w)=0\}\subseteq W, \\
  \im{f(v)} &= \{w'\in W'\colon w'=f(v)(w), \ w\in W\}\subseteq W'.
\end{align*}
Associated with the map $f(v)$ is the transpose map
\begin{align*}
  f'(v)\colon W'\to W, \quad f'(v)(w')=v\otimes w^{\prime *}.
\end{align*}
The matrices of $f(v)$ and $f'(v)$ are the transposes of each other.

Introducing inner products in $W$ and $W'$, we can relate the kernels and images of $f(v)$ and $f'(v)$ through orthogonal compliments,
\begin{align*}
  \im{f(v)}=(\ker{f'(v)})^\perp, \quad \im{f'(v)}=(\ker{f(v)})^\perp.
\end{align*}
(The orthogonal complement $Y^\perp$ of a subspace $Y$ of an inner product space $X$ is the set of all vectors in $X$ that are orthogonal to every vector in $Y$, $Y^\perp=\bigl\{x\in X\colon\langle x,y\rangle=0, \ \forall y\in Y \bigr\}$.)

Thus, if both maps are used to construct $L(v)$, then it suffices to consider only their kernels, for example.
We adopt this choice.
Furthermore, since $f'(v)$ is obtained from $f(v)$ by interchanging $W$ and $W'$, both maps are included by considering only $f(v)$ for both $V=W\otimes W'$ and $V=W'\otimes W$.

For specific computations we need expressions for the above quantities in terms of coordinates.
We introduce these by choosing arbitrary bases $\{e_i\}_{1\le i\le\dim{W}}$ and $\{e'_i\}_{1\le i\le\dim{W'}}$ for the spaces $W$ and $W'$ and representing a vector $v\in V$ in terms of its coordinates $\{v_{i,j}\}$,
\begin{align*}
  v=\sum_{i=1}^{\dim{W}} \sum_{j=1}^{\dim{W'}} v_{i,j} e_i\otimes e'_j.
\end{align*}
We find
\begin{align*}
  f(v)(w)&=\sum_{i=1}^{\dim{W}} \sum_{j=1}^{\dim{W'}} v_{i,j} w_i e'_j,\\
  \ker{f(v)} &= \Bigl\{w\in W\colon \sum_{i=1}^{\dim{W}} v_{i,j} w_i=0, \ j\in\{1,\ldots,\dim{W'}\}\Bigr\}.
\end{align*}
The kernel of a map is found by solving a homogeneous system of linear equations.

\subsection*{Partitions}

To describe properties of $v$ related to partitioning the system into any two subsystems, we need to consider all possible subsystems $T$ and $T'$ such that $S=T\cup T'$ and the corresponding $W$ and $W'$ such that $V=W\otimes W'$.
These quantities are given by
\begin{align*}
  T=S_J, \quad T'=S_{I\setminus J}, \quad W=V_J, \quad W'=V_{I\setminus J}, \quad J\in P'(I), \\
  S_H=\cup_{h\in H}S_h, \quad V_H=\otimes_{h\in H}V_h, \quad H\in P'(I).
\end{align*}
Here $I\setminus J$ is the relative complement of $J$ in $I$.
Also, $P'(I)=P(I)\setminus\{\varnothing,I\}$, where $P(I)$ is the power set of $I$ (the set of all subsets of $I$).
We use $P'(I)$ instead of $P(I)$ to exclude partitions with empty subsystems $(T,T')=(\varnothing,S)$ and $(T,T')=(S,\varnothing)$.

Now, for each $J\in P'(I)$, we define the corresponding map
\begin{align*}
  f_J(v)\colon V_J\to V_{I\setminus J}, \quad f_J(v)(w)=v\otimes w^*,
\end{align*}
its kernel $K_J(v)=\ker{f_J(v)}$, and its nullity $n_J(v)=\dim{K_J(v)}$.

In terms of arbitrary bases $\{e_i\}_{1\le i\le\dim{V_J}}$ and $\{e'_i\}_{1\le i\le\dim{V_{I\setminus J}}}$ for the spaces $V_J$ and $V_{I\setminus J}$, we have
\begin{align*}
  v&=\sum_{i=1}^{\dim{V_J}} \sum_{j=1}^{\dim{V_{I\setminus J}}} v_{i,j} e_i\otimes e'_j,\\
  f_J(v)(w)&=\sum_{i=1}^{\dim{V_J}} \sum_{j=1}^{\dim{V_{I\setminus J}}} v_{i,j} w_i e'_j,\\
  K_J(v)&= \Bigl\{w\in V_J\colon \sum_{i=1}^{\dim{V_J}} v_{i,j} w_i=0, \ j\in\{1,\ldots,\dim{V_{I\setminus J}}\}\Bigr\}.
\end{align*}

\subsection*{Example of three qubits}

For three qubits we have
\begin{align*}
  I=\{1,2,3\}, \quad P'(I)=\{\{1\},\{2\},\{3\},\{1, 2\},\{1, 3\},\{2, 3\}\}, 
\end{align*}
which gives the maps
\begin{align*}
  f_{\{1\}}(v)\colon V_1\to V_2\otimes V_3, \quad f_{\{1\}}(v)(w)&=\sum_{j_1=1}^2\sum_{j_2=1}^2\sum_{j_3=1}^2 v_{j_1,j_2,j_3}w_{j_1} e_{2,j_2}\otimes e_{3,j_3},\\ 
  f_{\{2\}}(v)\colon V_2\to V_1\otimes V_3, \quad f_{\{2\}}(v)(w)&=\sum_{j_1=1}^2\sum_{j_2=1}^2\sum_{j_3=1}^2 v_{j_1,j_2,j_3}w_{j_2} e_{1,j_1}\otimes e_{3,j_3},\\ 
  f_{\{3\}}(v)\colon V_3\to V_1\otimes V_2, \quad                            f_{\{3\}}(v)(w)&=\sum_{j_1=1}^2\sum_{j_2=1}^2\sum_{j_3=1}^2 v_{j_1,j_2,j_3}w_{j_3} e_{1,j_1}\otimes e_{2,j_2},\\ 
  f_{\{1,2\}}\colon V_1\otimes V_2\to V_3, \quad f_{\{1,2\}}(v)(w)&=\sum_{j_1=1}^2\sum_{j_2=1}^2\sum_{j_3=1}^2 v_{j_1,j_2,j_3}w_{j_1,j_2}e_{3,j_3},\\ 
  f_{\{1,3\}}\colon V_1\otimes V_3\to V_2, \quad f_{\{1,3\}}(v)(w)&=\sum_{j_1=1}^2\sum_{j_2=1}^2\sum_{j_3=1}^2 v_{j_1,j_2,j_3}w_{j_1,j_3}e_{2,j_2},\\ 
  f_{\{2,3\}}\colon V_2\otimes V_3\to V_1, \quad f_{\{2,3\}}(v)(w)&=\sum_{j_1=1}^2\sum_{j_2=1}^2\sum_{j_3=1}^2 v_{j_1,j_2,j_3}w_{j_2,j_3}e_{1,j_1}.
\end{align*}

\subsection*{Tensor products of maps}

To obtain the complete entanglement information about a vector $v$, we need to describe its properties related to partitioning the system $S$ into any number of subsystems.
For this purpose we construct the set of new maps $\{\tilde{f}_J(v)\}_{J\in P'(I)}$ from the set  of the maps $\{f_J(v)\}_{J\in P'(I)}$ using the operation of the tensor product.
The new maps should be linear in $v$, and it should be possible to compare them with each other, for example, by comparing their kernels.
Linearity in $v$ requires that the only other maps allowed in the construction are the identity maps.
Comparison of the new maps is possible only if their domains coincide, and a natural choice for such a common domain is the space $V$.
These requirements fix the form of the new maps,
\begin{align*}
  \tilde{f}_J(v)\colon V\to V_{I\setminus J}\otimes V_{I\setminus J}, \quad \tilde{f}_J(v)=f_J(v)\otimes\id_{I\setminus J}, \quad \tilde{f}_J(v)(x\otimes y)=(v\otimes x^*)\otimes y,
\end{align*}
where $\id_{W'}\colon W'\to W'$ is the identity map. 
Let $\tilde{K}_J(v)=\ker{\tilde{f}_J(v)}$ and $\tilde{n}_J(v)=\dim{\tilde{K}_J(v)}$ for each $J\in P'(I)$.
We note the relation $\tilde{K}_J(v)=K_J(v)\otimes V_{I\setminus J}$, which follows from the identities
\begin{align*}
  \ker{(f_J(v)\otimes\id_{I\setminus J})}=\ker{f_J(v)}\otimes V_{I\setminus J}+V_J\otimes\ker{\id_{I\setminus J}}
\end{align*}
and $\ker{\id_{I\setminus J}}=\{0\}$.

In terms of arbitrary bases $\{e_i\}_{1\le i\le\dim{V_J}}$ and $\{e'_i\}_{1\le i\le\dim{V_{I\setminus J}}}$ for the spaces $V_J$ and $V_{I\setminus J}$, we have
\begin{align*}
  v&=\sum_{i=1}^{\dim{V_J}} \sum_{j=1}^{\dim{V_{I\setminus J}}} v_{i,j} e_i\otimes e'_j,\\
  \tilde{f}_J(v)(w)&=\sum_{i=1}^{\dim{V_J}} \sum_{j=1}^{\dim{V_{I\setminus J}}} \sum_{k=1}^{\dim{V_{I\setminus J}}} v_{i,j} w_{i,k} e'_j\otimes e'_k,\\
  \tilde{K}_J(v)&= \Bigl\{w\in V_J\colon \sum_{i=1}^{\dim{V_J}} v_{i,j} w_{i,k}=0, \ j\in\{1,\ldots,\dim{V_{I\setminus J}}\}, \ k\in\{1,\ldots,\dim{V_{I\setminus J}}\}    \Bigr\}.
\end{align*}

Finally, the set $L(v)=\{\tilde{K}_J(v)\}_{J\in P'(I)}$ is the desired set of subspaces of $V$ that describes entanglement properties of $v$.

\subsection*{Example of three qubits}

For three qubits the maps $\tilde{f}_J(v)$ are
\begin{align*}
  &\tilde{f}_{\{1\}}(v)\colon V\to V_2\otimes V_3\otimes V_2\otimes V_3,\\
  &\tilde{f}_{\{1\}}(v)(w)=\sum_{j_1=1}^2\sum_{j_2=1}^2\sum_{j_3=1}^2\sum_{k_2=1}^2\sum_{k_3=1}^2  v_{j_1,j_2,j_3}w_{j_1,k_2,k_3} e_{2,j_2}\otimes e_{3,j_3}\otimes e_{2,k_2}\otimes e_{3,k_3},\\ 
  &\tilde{f}_{\{2\}}(v)\colon V\to V_1\otimes V_3\otimes V_1\otimes V_3,\\
  &\tilde{f}_{\{2\}}(v)(w)=\sum_{j_1=1}^2\sum_{j_2=1}^2\sum_{j_3=1}^2\sum_{k_1=1}^2\sum_{k_3=1}^2  v_{j_1,j_2,j_3}w_{k_1,j_2,k_3} e_{1,j_1}\otimes e_{3,j_3}\otimes e_{1,k_1}\otimes e_{3,k_3},\\ 
  &\tilde{f}_{\{3\}}(v)\colon V\to V_1\otimes V_2\otimes V_1\otimes V_2,\\
  &\tilde{f}_{\{3\}}(v)(w)=\sum_{j_1=1}^2\sum_{j_2=1}^2\sum_{j_3=1}^2\sum_{k_1=1}^2\sum_{k_2=1}^2  v_{j_1,j_2,j_3}w_{k_1,k_2,j_3} e_{1,j_1}\otimes e_{2,j_2}\otimes e_{1,k_1}\otimes e_{2,k_2},\\ 
  &\tilde{f}_{\{1,2\}}(v)\colon V\to V_3\otimes V_3,\\
  &\tilde{f}_{\{1,2\}}(v)(w)=\sum_{j_1=1}^2\sum_{j_2=1}^2\sum_{j_3=1}^2\sum_{k_3=1}^2 v_{j_1,j_2,j_3}w_{j_1,j_2,k_3}e_{3,j_3}\otimes e_{3,k_3},\\ 
  &\tilde{f}_{\{1,3\}}(v)\colon V\to V_2\otimes V_2,\\
  &\tilde{f}_{\{1,3\}}(v)(w)=\sum_{j_1=1}^2\sum_{j_2=1}^2\sum_{j_3=1}^2\sum_{k_2=1}^2 v_{j_1,j_2,j_3}w_{j_1,k_2,j_3}e_{2,j_2}\otimes e_{2,k_2},\\ 
  &\tilde{f}_{\{2,3\}}(v)\colon V\to V_1\otimes V_1,\\
  &\tilde{f}_{\{2,3\}}(v)(w)=\sum_{j_1=1}^2\sum_{j_2=1}^2\sum_{j_3=1}^2\sum_{k_1=1}^2 v_{j_1,j_2,j_3}w_{k_1,j_2,j_3}e_{1,j_1}\otimes e_{1,k_1}. 
\end{align*}

\subsection*{Spaces and their intersections}

Results in linear algebra \cite{algebra} show that the complete information about a set of linear subspaces is given by the dimensions of the subspaces and of all their intersections.
Each linear space is identified by its dimension, and the intersections are needed to account for the relative positions of the subspaces.
We specify such intersections for each set of subsets of $I$,
\begin{align*}
  \tilde{K}_Q(v)=\cap_{J\in Q}\tilde{K}_J(v), \quad \tilde{n}_Q(v)=\dim{\tilde{K}_Q(v)}, \quad Q\in P(P'(I)).
\end{align*}
Consequently, considering all such intersections, we find the sequence (ordered set) of new invariants describing all entanglement properties of $v$,
\begin{align*}
  \tilde{N}(v)=(\tilde{n}_Q(v))_{Q\in P(P'(I))}.
\end{align*}
We order elements of $\tilde{N}(v)$ using the canonical ordering of elements of $P(I)$, which is obtained from binary representations of elements of $P(I)$ considered as $\{0,1\}^I$.
We call elements of $\tilde{N}(v)$ algebraic invariants of $v$ because they are derived using standard tools of linear algebra. 

Finally, we define the equivalence relation $\sim''_{V}$ on the set $V$ by setting $v'\sim''_{V}v$ if and only if $\tilde{N}(v')=\tilde{N}(v)$, for each $v,v'\in V$.
The relation $\sim_{V}''$ defines the quantities $C''(v)$, $\tilde{v}''$, $C''$, $\tilde{V}''$ in the same manner as the relations $\sim_{V}$ and $\sim'_{V}$ define the quantities $C(v)$, $\tilde{v}$, $C$, $\tilde{V}$ and $C'(v)$, $\tilde{v}'$, $C'$, $\tilde{V}'$, respectively.

The proceeding development shows that the equivalence relations $\sim'_V$ and $\sim''_V$ are identical and proves the following theorem.
\begin{theorem}\label{theorem}
  There is a one-to-one correspondence between the quotient set $C'$ and the sequence of values of the algebraic invariants $(\tilde{N}(v))_{v\in V}$.
\end{theorem}

\subsection*{Independent invariants}

In general, there are certain algebraic relations between elements of $\tilde{N}(v)$.
For example,
\begin{align*}
  \dim{V_J}-\frac{\tilde{n}_J(v)}{\dim{V_{I\setminus J}}}=\dim{V_{I\setminus J}}-\frac{\tilde{n}_{I\setminus J}(v)}{\dim{V_J}}
\end{align*}
is true for all $v\in V$ and $J\in P'(I)$. 
We can say, for example, that $\tilde {n}_J(v)$ is an independent invariant and $\tilde {n}_{I\setminus J}(v)$ is a dependent invariant, which can be done consistently by taking $J$ only from an appropriate subset of $P'(I)$.

It is convenient to remove dependent elements from $\tilde{N}(v)$ by defining a subsequence of independent invariants
\begin{align*}
  \tilde{N}'(v)=(\tilde{n}_Q(v))_{Q\in R}, \quad R\subseteq P(P'(I)),
\end{align*}
which we call a generating sequence of invariants of $v$.
For consistency, we use the same $R$ for each $v\in V$.
We order elements of $\tilde{N}'(v)$ canonically.
All elements of $\tilde{N}'(v)$ are algebraically independent of each other, and all elements of $\tilde{N}(v)$ which are not in $\tilde{N}'(v)$ can be algebraically expressed in terms of elements of $\tilde{N}'(v)$.
We choose $R$ with the smallest number of elements; although this choice is not unique, all such choices are equivalent for our purposes.

It remains to choose the set $R$.
We define $R=\lim_{k\to\infty}R_k$, where the sequence of sets $(R_1,R_2,\dotsc)$ is such that $R_k\supseteq R_{k+1}$ for each $k\in\N$.
We set $R_1=P(P'(I))$ and find the elements of the sequence iteratively by the following steps that remove dependent invariants:
\begin{enumerate}
  \item If there exist $X\in R'_k$ and $X_1,X_2\in X$ such that $X_1\subseteq X_2$, then $R''_k=(R'_k\setminus\{X\})\cup(X\setminus\{X_1\})$ for any such $X,X_1$; otherwise, $R''_k=R'_k$.
  \item If there exists $X\in R''_k$ such that $X_1\cap X_2=\varnothing$ for any $X_1,X_2\in X$, then $R'''_k=R''_k\setminus\{X\}$ for any such $X$; otherwise, $R'''_k=R''_k$.
  \item If there exist $X,Y\in R'''_k$ such that $X_1\in Y$ for any $X_1\in X$, then $R_{k+1}=R'''_k\setminus\{X\}$ for any such $X$; otherwise, $R_{k+1}=R'''_k$.
\end{enumerate}
If there is more than one choice for $X$ (and for $X_1$ in step $1$) that satisfies the conditions in a given step, then any such choice can be made.
(The resulting sequence $(R_1,R_2,\dotsc)$ depends on these choices.)
For any such choice, however, the sequence is convergent and its limit $R=\lim_{k\to\infty}R_k$ is reached after a finite number of iterations, i.e. there exists $m\in\N$ such that $R_k=R$ for each $k\ge m$.
Even though the above choices can lead to different sets $R$ and resulting generating sequences $\tilde{N}'(v)$, they lead to the same entanglement classification.
This completes the construction of each generating sequence of invariants $\tilde{N}'(v)$.

The relation
\begin{align*}
  \tilde{K}_Q(v)=\cap_{J\in Q}(K_J(v)\otimes V_{I\setminus J}), \quad Q\in P(P'(I)) 
\end{align*}
implies 
\begin{align*}
  \tilde{n}_Q(v)=n_Q(v)\dim{V_{I\setminus\cup_{J\in Q}J}}, \quad \cup_{J\in Q}J\subset I, \quad Q\in P(P'(I)), 
\end{align*}
where we set $\dim{V_{\varnothing}}=1$.
(To prove this, note that $\tilde{K}_Q(v)$ does not involve $K_{J'}(v)$, where $J'\in I\setminus\cup_{J\in Q}J$.)
For $n\ge 3$, this relation between the invariants means that such $\tilde{n}_Q(v)$ describes properties of $v$ related to partitioning the system into at most $\abs{I\backslash\cup_{J\in Q}J}$ subsystems.
For such cases, it is convenient to replace $\tilde{n}_Q(v)$ with $n_Q(v)$ and define the set of invariants
\begin{align*}
  \tilde{N}''(v)=(m_Q(v))_{Q\in R}, \quad m_Q(v)=
  \begin{cases}
    n_Q(v), & \cup_{J\in Q}J\subsetneq I,\\
    \tilde{n}_Q(v), & \cup_{J\in Q}J=I.
  \end{cases}
\end{align*}
We order elements of $\tilde{N}''(v)$ canonically.
We give our explicit solutions in terms of $\tilde{N}''(v)$.

\subsection*{General forms of states}

As our main computational device, we use the general forms of elements of $\tilde{V}''$.
We obtain them from expressions for elements of $\ker{f(v)}$ for a map $f(v)\colon W\to W'$, to derivation of which we now turn.
We choose arbitrary bases $\{u_i\}_{1\le i\le\dim{W}}$ and $\{u'_i\}_{1\le i\le\dim{W'}}$ for the spaces $W$ and $W'$, respectively, and represent a vector $v\in V$ in terms of its coordinates,
\begin{align*}
  v=\sum_{i=1}^{\dim{W}} \sum_{j=1}^{\dim{W'}} v_{i,j} u_i\otimes u'_j, \quad \{v_{i,j}\}\subset F.
\end{align*}
It follows that $v$ decomposes according to
\begin{align*}
  v&=\sum_{i=1}^{\dim{W}} u_i\otimes \tilde{u}'_i, \quad \tilde{u}'_i=\sum_{j=1}^{\dim{W'}}v_{i,j}u'_j, \quad \{\tilde{u}'_i\}\subset W', \\
  v&=\sum_{j=1}^{\dim{W'}} \tilde{u}_j\otimes u'_j, \quad \tilde{u}_j=\sum_{i=1}^{\dim{W}}v_{i,j}u_i, \quad \{\tilde{u}_j\}\subset W.
\end{align*}
The defining relation $v\otimes w^*=0$ for $w\in\ker{f(v)}$, which is a system of homogeneous linear equations for the coordinates of $w$, now implies the general form of $v$,
\begin{align*}
  v=\sum_{i=1}^{\dim{W}-n(v)} w_i\otimes w'_i, \quad \{w_i\}\subset W, \quad \{w'_i\}\subset W', \\
  \dim{\linearspan{(\{w_i\})}}=\dim{\linearspan{(\{w'_i\})}}=\dim{W}-n(v),
\end{align*}
where $n(v)=\dim{\ker{f(v)}}$ and the dimension of the span of a set of vectors is the number of its linearly independent elements.
This decomposition is unique up to linear transformations $w_i\mapsto\sum_j B_{i,j}w_j$ and $w'_i\mapsto\sum_j B'_{i,j}w'_j$, where $B$ and $B'$ are nonsingular square matrices of order $\dim{W}-n(v)$ that satisfy the condition $B^t B'=1$.

When considering the above general forms of elements of $V$ resulting from different choices of $W$ and $W'$ such that $V=W\otimes W'$, we need to choose $\{w_i\}$ and $\{w'_i\}$ (using appropriate $B$ and $B'$) such that the corresponding decompositions are consistent for all such choices. 
This results in restrictions on allowed values of the invariants in $\tilde{N}(v)$ and, consequently, leads to the classification of all entangled states.

The described method solves the restricted classification problem for arbitrary $\{V_i\}_{i\in I}$.
Obtaining explicit solutions, however, is entirely different matter.
We did not obtain such solutions for arbitrary $\{V_i\}_{i\in I}$, but we found them for numerous examples given in the following section.

Particularly interesting are cases where the spaces in $\{V_i\}_{i\in I}$ are of equal dimensions.
The resulting permutation symmetry among the spaces reduces the equivalence classes to sets of classes related by the symmetry.
As a result, representative elements for the sets of classes take simple forms.
We have explicit solutions for two such symmetric examples. 

\section{Examples}

The present classification method works for arbitrary finite $n$ and $D=(\dim{V_i})_{i\in I}$.
The case $n=2$ is easily solved~\cite{Buniy:2010yh} for arbitrary $D$.
We now apply our method to the case $n=3$ for a large selection of values of $D$ and the case $n=4$, $D=(2,2,2,2)$ (four qubits).

\subsection{$\bm{n=3}$}

Independent invariants for $n=3$ are given by the sets
\begin{align*}
  Q_1=\{\{1\}\}, \quad Q_2=\{\{2\}\}, \quad Q_3=\{\{3\}\}, \quad Q_4=\{\{1,2\},\{1,3\},\{2,3\}\}.
\end{align*}
The sets $Q_1,Q_2,Q_3$ and $Q_4$ lead to invariants related to partitioning the system into two and three subsystems, respectively.
For each of these invariants, there are corresponding invariants generated by the transpose maps, which do not need to be considered.
Since all other partitions lead to dependent invariants, we choose the generating set of invariants
\begin{align*}
  \tilde{N}''(v)=(n_{Q_1}(v),n_{Q_2}(v),n_{Q_3}(v),\tilde{n}_{Q_4}(v))
\end{align*}
for each $v\in V$.

For the set of equivalent classes we find
\begin{align*}
  C''=\{C_0\}\cup\{C_{k_1,k_2,k_3,j}\colon k_1\in\{1,\dotsc,d_1\}, k_2\in\{1,\dotsc,d_2\}, k_3\in\{1,\dotsc,d_3\}, j\in M_{k_1,k_2,k_3}\},
\end{align*}
where $D=(d_1,d_2,d_3)$ and $M_{k_1,k_2,k_3}$ is a certain set of natural numbers that is symmetric in $k_1,k_2,k_3$.
The values of the invariants in $\tilde{N}''(v)$ for the classes $C_0$ and $C_{k_1,k_2,k_3,j}$ are given in Table \ref{table:invariants}.
\begin{table}[ht]
  \caption{\label{table:invariants} The values of the invariants in $\tilde{N}''(v)$ for $n=3$, $D=(d_1,d_2,d_3)$.}
\footnotesize
  \begin{ruledtabular}
    \begin{tabular}{ccccc}
      & $n_{Q_1}(v)$ & $n_{Q_2}(v)$ & $n_{Q_3}(v)$ & $\tilde{n}_{Q_4}(v)$ \\ \hline
      $C_0$ & $d_1$ & $d_2$ & $d_3$ & $d_1 d_2 d_3$ \\
      $C_{k_1,k_2,k_3,j}$ & $d_1-k_1$ & $d_2-k_2$ & $d_3-k_3$ & $d_1 d_2 d_3 -k_1 d_1 -k_2 d_2 -k_3 d_3 +(M_{k_1,k_2,k_3})_j$ \\
    \end{tabular}
  \end{ruledtabular}
\end{table}
Although we do not have a general formula for $M_{k_1,k_2,k_3}$ for arbitrary $(k_1,k_2,k_3)$, we give $M_{k_1,k_2,k_3}$ for various particular values of $(k_1,k_2,k_3)$ in Table \ref{table:m}, which is our main result for the case $n=3$.
With analogous computations for additional values of $(k_1,k_2,k_3)$, the table can be easily expanded.
Such a table is directly used for explicit computations of $C''$ for various values of $D$.
In particular, the values of $M_{k_1,k_2,k_3}$ given in Table \ref{table:m} suffice to find the set of classes $C''$ for each value of $D$ given in Table \ref{table:number-of-classes}; the latter table gives only the number of classes $\abs{C''}$. 
As illustrative examples and because of space limits, we present here the full results only for $D=(2,2,d)$ and $D=(2,3,d)$, where $d$ is arbitrary, in Tables \ref{table:22d} and \ref{table:23d}, respectively.
For the symmetric case $D=(2,2,2)$, there are $5$ sets of classes related by permutations of $\{V_1,V_2,V_3\}$; Table \ref{table:222-permutations} lists the sets and their representative elements.
\begin{table}[ht]
  \caption{\label{table:m} The set $M_{k_1,k_2,k_3}$ for various values of $(k_1,k_2,k_3)$.
  The notation $(m,\dotsc,m')$ means all integers between and including $m$ and $m'$.}
\footnotesize
  \begin{ruledtabular}
    \begin{tabular}{llp{10ex}ll}
      $(k_1,k_2,k_3)$ & $M_{k_1,k_2,k_3}$ & & $(k_1,k_2,k_3)$ & $M_{k_1,k_2,k_3}$ \\
      \cline{1-2}                                      
      \cline{4-5}                                      
      $(1,1,1)$ & $(2)$ & &                            $(2,6,6)$ & $(7,\dotsc,23,28,29)$ \\
      $(1,2,2)$ & $(5)$ & &                            $(2,6,7)$ & $(5,\dotsc,22,24,25,26,34)$ \\
      $(1,3,3)$ & $(10)$ & &                           $(2,6,8)$ & $(8,\dotsc,25,31)$ \\
      $(1,4,4)$ & $(17)$ & &                           $(2,6,9)$ & $(13,14,16,\dotsc,20,22,\dotsc,26,29,30)$ \\
      $(1,5,5)$ & $(26)$ & &                           $(2,6,10)$ & $(20,23,24,28,29,31)$ \\
      $(1,6,6)$ & $(37)$ & &                           $(2,6,11)$ & $(29,34)$ \\
      \cline{1-2}                                      
      $(2,2,2)$ & $(4,5)$ & &                          $(2,6,12)$ & $(40)$ \\
      \cline{4-5}                                      
      $(2,2,3)$ & $(5,6)$ & &                          $(3,3,3)$ & $(2,\dotsc,8,10)$ \\                                                            
      $(2,2,4)$ & $(8)$ & &                            $(3,3,4)$ & $(2,\dotsc,11)$ \\
      \cline{1-2}                                      
      $(2,3,3)$ & $(4,\dotsc,8)$ & &                   $(3,3,5)$ & $(2,\dotsc,11,14)$ \\
      $(2,3,4)$ & $(5,\dotsc,8,10)$ & &                $(3,3,6)$ & $(2,\dotsc,12)$ \\
      $(2,3,5)$ & $(8,10)$ & &                         $(3,3,7)$ & $(4,\dotsc,11,14)$ \\
      $(2,3,6)$ & $(13)$ & &                           $(3,3,8)$ & $(10,11,14)$ \\                                                            
      \cline{1-2}                                      
      $(2,4,4)$ & $(5,\dotsc,13)$ & &                  $(3,3,9)$ & $(18)$ \\
      \cline{4-5}
      $(2,4,5)$ & $(5,\dotsc,13,16)$ & &               $(3,4,4)$ & $(2,\dotsc,12,14)$ \\                                                                
      $(2,4,6)$ & $(8,9,10,12,13,15)$ & &              $(3,4,5)$ & $(2,\dotsc,16)$ \\
      $(2,4,7)$ & $(13,16)$ & &                        $(3,4,6)$ & $(2,\dotsc,16,20)$ \\                                                            
      $(2,4,8)$ & $(20)$ & &                           $(3,4,7)$ & $(2,\dotsc,17)$ \\
      \cline{1-2}                                      
       $(2,5,5)$ & $(6,\dotsc,16,19,20)$ & &           $(3,4,8)$ & $(2,\dotsc,17,19)$ \\                                                            
       $(2,5,6)$ & $(5,\dotsc,18,24)$ & &              $(3,4,9)$ & $(2,\dotsc,17,20)$ \\
       $(2,5,7)$ & $(8,10,\dotsc,18,20,22)$ & &        $(3,4,10)$ & $(5,\dotsc,16,19,20)$ \\
       $(2,5,8)$ & $(13,15,16,19,20,22)$ & &           $(3,4,11)$ & $(14,16,20)$ \\
       $(2,5,9)$ & $(20,24)$ & &                       $(3,4,12)$ & $(25)$ \\
       $(2,5,10)$ & $(29)$ & &                         
    \end{tabular}
  \end{ruledtabular}
\end{table}
\begin{table}[ht]
  \caption{\label{table:number-of-classes} The numbers of equivalence classes $\abs{C''}$ for $n=3$ and various values of $D$.}
\footnotesize
  \begin{ruledtabular}
    \begin{tabular}{lcp{10ex}lcp{10ex}lc}
      $D$ & $\abs{C''}$ & & $D$ & $\abs{C''}$ & & $D$ & $\abs{C''}$ \\
      \cline{1-2}
      \cline{4-5}
      \cline{7-8}
      $(2,2,2)$         & $7$ & &    $(2,5,5)$          & $77$ & &    $(3,3,3)$         & $39$ \\  
      $(2,2,3)$         & $9$ & &    $(2,5,6)$          & $99$ & &    $(3,3,4)$         & $60$ \\  
      $(2,2,d),d\ge 4$  & $10$ & &   $(2,5,7)$          & $113$ & &   $(3,3,5)$         & $75$ \\ 
      \cline{1-2}
      $(2,3,3)$         & $17$ & &   $(2,5,8)$          & $120$ & &   $(3,3,6)$         & $88$ \\ 
      $(2,3,4)$         & $23$ & &   $(2,5,9)$          & $122$ & &   $(3,3,7)$         & $97$ \\ 
      $(2,3,5)$         & $25$ & &   $(2,5,d), d\ge 10$ & $123$ & &   $(3,3,8)$         & $100$ \\ 
      \cline{4-5}
      $(2,3,d),d\ge 6$  & $26$ & &   $(2,6,6)$          & $141$ & &   $(3,3,d), d\ge 9$ & $101$ \\ 
      \cline{1-2}                 
      \cline{7-8}                 
      $(2,4,4)$         & $39$ & &   $(2,6,7)$          & $177$ & &   $(3,4,4)$          & $103$ \\
      $(2,4,5)$         & $51$ & &   $(2,6,8)$          & $203$ & &   $(3,4,5)$          & $143$ \\
      $(2,4,6)$         & $58$ & &   $(2,6,9)$          & $219$ & &   $(3,4,6)$          & $178$ \\
      $(2,4,7)$         & $60$ & &   $(2,6,10)$         & $226$ & &   $(3,4,7)$          & $205$ \\
      $(2,4,d), d\ge 8$ & $61$ & &   $(2,6,11)$         & $228$ & &   $(3,4,8)$          & $226$ \\
      & & &                          $(2,6,d), d\ge 12$ & $229$ & &   $(3,4,9)$          & $244$ \\                               
      & & &                          & & &                            $(3,4,10)$         & $258$ \\
      & & &                          & & &                            $(3,4,11)$         & $261$ \\
      & & &                          & & &                            $(3,4,d), d\ge 12$ & $262$ \\
    \end{tabular}                                                             
  \end{ruledtabular}
\end{table}
\begin{table}[ht]
  \caption{\label{table:22d} The entanglement classes, their algebraic invariants, and their representative elements for $n=3$, $D=(2,2,d)$.
  Classes for which any of the invariants in the set $\tilde{N}''(v)$ are negative should be discarded.
  Classes within a horizontal block are added each time $d$ increases by $1$, so that there are $7$, $9$, $10$ classes for $d=2$, $d=3$, $d\ge 4$, respectively.
  Each expression $[j_1,j_2,j_3]$ stands for $u_{1,j_1}\otimes u_{2,j_2}\otimes u_{3,j_3}$, where $\{u_{i,j}\}$ is a set of any linearly independent elements of $V_i$.}
\footnotesize
  \begin{ruledtabular}
    \begin{tabular}{lll}
            & $\tilde{N}''(v)$           & $v$ \\ \hline
      $C_0$ & $(2,2,d,4d)$     & $0$ \\ \hline
      $C_1$ & $(1,1,d-1,3d-2)$ & $[1,1,1]$ \\
      $C_2$ & $(0,0,d-1,3d-3)$ & $[1,1,1]+[2,2,1]$ \\ \hline
      $C_3$ & $(0,1,d-2,2d-1)$ & $[1,1,1]+[2,1,2]$ \\
      $C_4$ & $(1,0,d-2,2d-1)$ & $[1,1,1]+[1,2,2]$ \\
      $C_5$ & $(0,0,d-2,2d-3)$ & $[1,1,1]+[1,2,2]+[2,1,2]$ \\
      $C_6$ & $(0,0,d-2,2d-4)$ & $[1,1,1]+[2,2,2]$ \\ \hline
      $C_7$ & $(0,0,d-3,d-2)$  & $[1,1,1]+[1,2,2]+[2,2,3]$ \\
      $C_8$ & $(0,0,d-3,d-3)$  & $[1,1,1]+[1,2,2]+[2,1,2]+[2,2,3]$ \\ \hline
      $C_9$ & $(0,0,d-4,0)$    & $[1,1,1]+[1,2,2]+[2,1,3]+[2,2,4]$ \\
    \end{tabular}
  \end{ruledtabular}
\end{table}
\begin{table}[ht]
  \caption{\label{table:23d} The entanglement classes, their algebraic invariants, and their representative elements for $n=3$, $D=(2,3,d)$.
  Classes for which any of the invariants in the set $\tilde{N}''(v)$ are negative should be discarded.
  Classes within a horizontal block are added each time $d$ increases by $1$, so that there are $9$, $17$, $23$, $25$, $26$ classes for $d=2$, $d=3$, $d=4$, $d=5$, $d\ge 6$, respectively.
  Each expression $[j_1,j_2,j_3]$ stands for $u_{1,j_1}\otimes u_{2,j_2}\otimes u_{3,j_3}$, where $\{u_{i,j}\}$ is a set of any linearly independent elements of $V_i$.}
\footnotesize
  \begin{ruledtabular}
    \begin{tabular}{lll}
               & $\tilde{N}''(v)$ & $v$ \\ \hline
      $C_0$    & $(2,3,d,6d)$     & $0$ \\ \hline
      $C_1$    & $(1,2,d-1,5d-3)$ & $[1,1,1]$ \\
      $C_2$    & $(0,1,d-1,5d-5)$ & $[1,1,1]+[2,2,1]$ \\ \hline
      $C_3$    & $(0,2,d-2,4d-2)$ & $[1,1,1]+[2,1,2]$ \\
      $C_4$    & $(1,1,d-2,4d-3)$ & $[1,1,1]+[1,2,2]$ \\
      $C_5$    & $(0,1,d-2,4d-5)$ & $[1,1,1]+[1,2,2]+[2,1,2]$ \\
      $C_6$    & $(0,1,d-2,4d-6)$ & $[1,1,1]+[2,2,2]$ \\
      $C_7$    & $(0,0,d-2,4d-7)$ & $[1,1,1]+[1,2,2]+[2,3,1]$ \\
      $C_8$    & $(0,0,d-2,4d-8)$ & $[1,1,1]+[1,2,2]+[2,2,1]+[2,3,2]$ \\ \hline
      $C_9$    & $(1,0,d-3,3d-1)$ & $[1,1,1]+[1,2,2]+[1,3,3]$ \\
      $C_{10}$ & $(0,1,d-3,3d-4)$ & $[1,1,1]+[1,2,2]+[2,1,3]$ \\
      $C_{11}$ & $(0,1,d-3,3d-5)$ & $[1,1,1]+[1,2,2]+[2,1,2]+[2,2,3]$ \\
      $C_{12}$ & $(0,0,d-3,3d-5)$ & $[1,1,1]+[1,2,2]+[1,3,3]+[2,1,2]$ \\
      $C_{13}$ & $(0,0,d-3,3d-6)$ & $[1,1,1]+[1,2,2]+[2,3,3]$ \\
      $C_{14}$ & $(0,0,d-3,3d-7)$ & $[1,1,1]+[1,2,2]+[1,3,3]+[2,1,2]+[2,2,3]$ \\
      $C_{15}$ & $(0,0,d-3,3d-8)$ & $[1,1,1]+[1,2,2]+[2,1,3]+[2,3,1]$ \\
      $C_{16}$ & $(0,0,d-3,3d-9)$ & $[1,1,1]+[1,2,2]+[2,2,2]+[2,3,3]$ \\ \hline
      $C_{17}$ & $(0,1,d-4,2d-2)$ & $[1,1,1]+[1,2,2]+[2,1,3]+[2,2,4]$ \\
      $C_{18}$ & $(0,0,d-4,2d-3)$ & $[1,1,1]+[1,2,2]+[1,3,3]+[2,3,4]$ \\
      $C_{19}$ & $(0,0,d-4,2d-5)$ & $[1,1,1]+[1,2,2]+[1,3,3]+[2,2,4]+[2,3,1]$ \\
      $C_{20}$ & $(0,0,d-4,2d-6)$ & $[1,1,1]+[1,2,2]+[2,2,3]+[2,3,4]$ \\
      $C_{21}$ & $(0,0,d-4,2d-7)$ & $[1,1,1]+[1,2,2]+[1,3,3]+[2,2,3]+[2,3,4]$ \\
      $C_{22}$ & $(0,0,d-4,2d-8)$ & $[1,1,1]+[1,2,2]+[1,3,3]+[2,1,2]+[2,2,3]+[2,3,4]$ \\ \hline
      $C_{23}$ & $(0,0,d-5,d-3)$  & $[1,1,1]+[1,2,2]+[1,3,3]+[2,1,4]+[2,2,5]$ \\
      $C_{24}$ & $(0,0,d-5,d-5)$  & $[1,1,1]+[1,2,2]+[1,3,3]+[2,1,3]+[2,2,4]+[2,3,5]$ \\ \hline
      $C_{25}$ & $(0,0,d-6,0)$    & $[1,1,1]+[1,2,2]+[1,3,3]+[2,1,4]+[2,2,5]+[2,3,6]$ \\
    \end{tabular}
  \end{ruledtabular}
\end{table}
\begin{table}[ht]
  \caption{\label{table:222-permutations} Representative elements for the sets of equivalence classes for $n=3$, $D=(2,2,2)$ induced by the permutation symmetry of the spaces in $\{V_1,V_2,V_3\}$.
  A representative element $v$ is given by $v=Av_1$, where $A\colon V\to V$ is a certain linear operator and $v_1\in V$ is a fixed vector. 
  (Without loss of generality and for comparison with other tables, we choose $v_1=[1,1,1]$.)
  The operator $a_i$ is defined by $a_i[\dotsc,1,\dotsc]=[\dotsc,2,\dotsc]$ and $a_i[\dotsc,2,\dotsc]=[\dotsc,1,\dotsc]$, where only the $i$th index changes.
  To obtain all classes in each group, all possible choices of the indices $\{i,j,k\}=\{1,2,3\}$ should be considered.}
\footnotesize
  \begin{ruledtabular}
    \begin{tabular}{ll}
      & $A$ \\ \hline
      $C_{0}$                 & $0$ \\
      $C_{1}$                 & $1$ \\
      $\{C_{2},C_{3},C_{4}\}$ & $1+a_i a_j$ \\
      $C_{5}$                 & $1+a_i(a_j+a_k)$ \\
      $C_{6}$                 & $1+a_i a_j a_k$ \\
    \end{tabular}
  \end{ruledtabular}
\end{table}

It is easy to obtain general expressions for $M_{k_1,k_2,k_3}$ for various particular values of $(k_1,k_2,k_3)$, and we give here just a few such results:
\begin{align*}
  M_{k_1,k_2,k_1 k_2} &= (k_1^2+k_2^2), \\
  M_{k_1,k_2,k_1 k_2-1} &= (\dotsc,k_1^2+k_2^2-2(k_1+k_2)+5,k_1^2+k_2^2-(k_1+k_2)+2).
\end{align*}
These and similar readily available expressions for $M_{k_1,k_2,k_3}$ suggest certain patterns, which might eventually lead to the general result for arbitrary $(k_1,k_2,k_3)$. 

\begin{figure}
  \includegraphics[width=260pt]{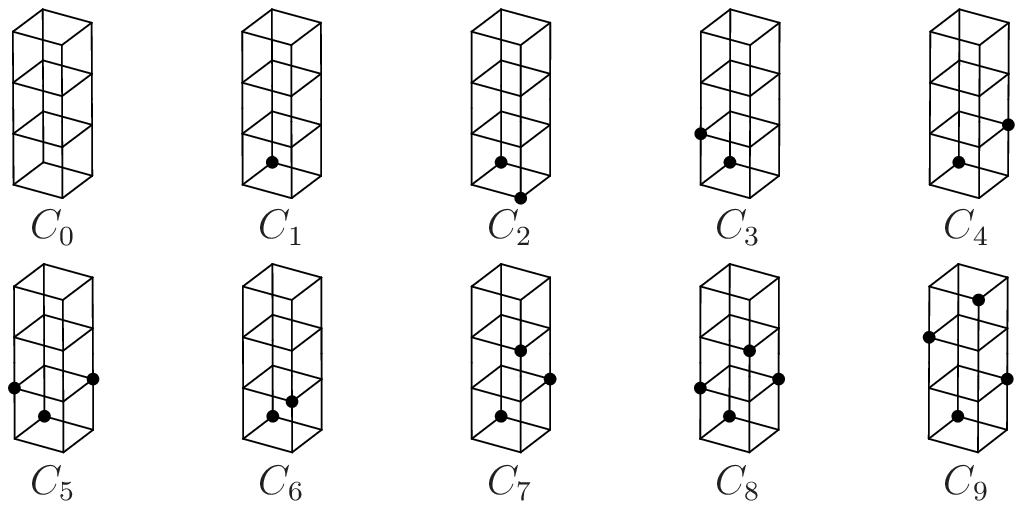}
  \caption{\label{figure:figure224} Graphical representation of entanglement classes for $n=3$, $D=(2,2,d)$.
  Each vertex corresponds to a certain expression $[j_1,j_2,j_3]$ in a representative element $v$ for each class, and $v$ is the sum of such expressions over all vertices of a given three-dimensional lattice; see Table~\ref{table:22d}.
  The invariants $n_{Q_1}(v)$, $n_{Q_2}(v)$, $n_{Q_3}(v)$ equal the numbers of linearly independent two-dimensional lattices.
  To obtain the corresponding representations for $D=(2,2,d)$, we remove $4-d$ cubes from tops of stacks  for $d<4$ or add $d-4$ cubes on top of stacks without adding any new vertices for $d>4$.
  There are only $10$ classes for any $d\ge 4$ because the number of linearly independent two-dimensional lattices along one of the directions is already maximal (four) for the class $C_9$.
  The construction for arbitrary $D$ is analogous; see, for example, Fig.~\ref{figure:figure236}.}
\end{figure}

\begin{figure}
  \includegraphics[width=440pt]{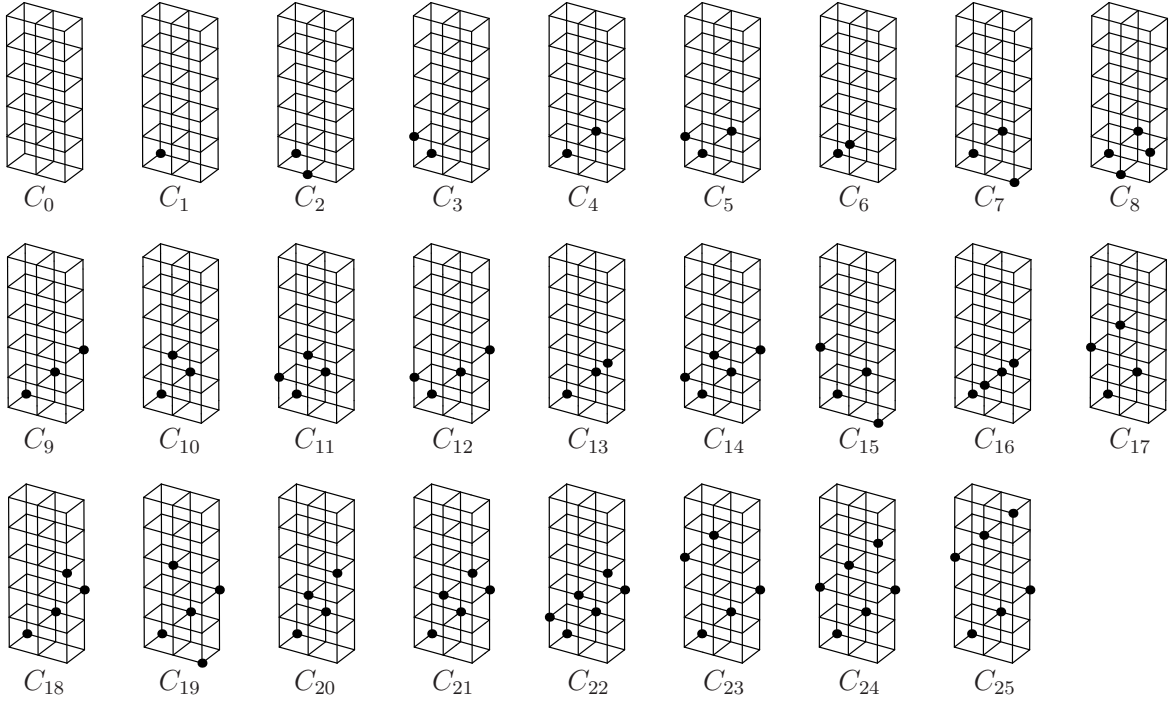}
  \caption{\label{figure:figure236} Graphical representation of entanglement classes for $n=3$, $D=(2,3,d)$.
  See Fig.~\ref{figure:figure224} for further details.}
\end{figure}

The needed computations for the above cases are lengthy but elementary, and we do not give their details here.
Instead, we invite the reader to study graphical representation of entanglement classes for the cases $D=(2,2,d)$ and $D=(2,3,d)$ in Figs.~\ref{figure:figure224} and \ref{figure:figure236}, respectively, which can be easily generalized for arbitrary $D$.
Although these and similar figures cannot replace the actual computations, they are useful in understanding relations between the classes, finding their general properties, and, perhaps, even in solving the general case.
In this regard, generalizations of Table \ref{table:222-permutations} seems to be particularly promising when solving the symmetric case $D=(d,d,d)$ for arbitrary $d$. 

\subsection{$\bm{n=4}$}

Table \ref{table-q4} lists sets that give independent invariants for $n=4$, arranged according to types of partitions of the system. 
\begin{table}[ht]
  \caption{\label{table-q4} Sets that give independent invariants for $n=4$.}
\footnotesize
  \begin{ruledtabular}
    \begin{tabular}{lp{2ex}l}
      $Q_1=\{\{1\}\}$ & & $Q_9=\{\{1,2\},\{1,3,4\},\{2,3,4\}\}$ \\
      $Q_2=\{\{2\}\}$ & & $Q_{10}=\{\{1,3\},\{1,2,4\},\{2,3,4\}\}$ \\
      $Q_3=\{\{3\}\}$ & & $Q_{11}=\{\{1,4\},\{1,2,3\},\{2,3,4\}\}$ \\
      $Q_4=\{\{4\}\}$ & & $Q_{12}=\{\{2,3\},\{1,2,4\},\{1,3,4\}\}$ \\
      \cline{1-1}
      $Q_5=\{\{1,2\},\{1,3\},\{2,3\}\}$ & & $Q_{13}=\{\{2,4\},\{1,2,3\},\{1,3,4\}\}$ \\
      $Q_6=\{\{1,2\},\{1,4\},\{2,4\}\}$ & & $Q_{14}=\{\{3,4\},\{1,2,3\},\{1,2,4\}\}$ \\
      \cline{3-3}
      $Q_7=\{\{1,3\},\{1,4\},\{3,4\}\}$ & & $Q_{15}=\{\{1,2\},\{1,3\},\{1,4\},\{2,3,4\}\}$ \\
      $Q_8=\{\{2,3\},\{2,4\},\{3,4\}\}$ & & $Q_{16}=\{\{1,2\},\{2,3\},\{2,4\},\{1,3,4\}\}$ \\
      & & $Q_{17}=\{\{1,3\},\{2,3\},\{3,4\},\{1,2,4\}\}$ \\
      & & $Q_{18}=\{\{1,4\},\{2,4\},\{3,4\},\{1,2,3\}\}$ \\
      \cline{3-3}
      & & $Q_{19}=\{\{1,2,3\},\{1,2,4\},\{1,3,4\},\{2,3,4\}\}$ \\
    \end{tabular}
  \end{ruledtabular}
\end{table}
The sets $Q_1,\dotsc,Q_4$ and $Q_5,\dotsc,Q_8$ lead to invariants related to partitioning the system into two and three subsystems, respectively.
Partitions into four subsystems are of three different types and are given by the sets $Q_9,\dotsc,Q_{14}$, and $Q_{15},\dotsc Q_{18}$, and $Q_{19}$.
For each of these five types, there are corresponding invariants generated by the transpose maps, which do not need to be considered.
Since all other partitions lead to dependent invariants, we choose the generating set of invariants
\begin{align*}
  \tilde{N}''(v)=(n_{Q_1}(v),\dotsc,n_{Q_8}(v),\tilde{n}_{Q_9}(v),\dotsc,\tilde{n}_{Q_{19}}(v))
\end{align*}
for each $v\in V$.

As an illustrative example, we take $D=(2,2,2,2)$.
It is convenient to partition $C''$ into three sets,
\begin{align*}
  C''=C''_{I,1}\cup C''_{I,2}\cup C''_{I,3},
\end{align*}
according to possible forms of representing elements for classes in each set.
The set $C''_{I,1}$ consists of classes that can be represented by elements with coefficients in linear combinations of bases vectors taken from $\{0,1\}$.
The set $C''_{I,2}$ consists of classes that do not belong to $C''_{I,1}$ and that can be represented by elements with coefficients in linear combinations of bases vectors taken from $\{0,1,-1\}$.
The set $C''_{I,3}$ consists of classes that do not belong to either $C''_{I,1}$ or $C''_{I,2}$.
The classes in $C''_{I,1}$ are the simplest and the most typical, and the classes in $C''_{I,3}$ are the most complex and the least typical.
It is clear that classes in $C''_{I,1}$ and $C''_{I,2}$ can be represented by elements with coefficients in linear combinations of bases vectors taken from other sets besides $\{0,1\}$ and other sets besides $\{0,1\}$, $\{0,1,-1\}$, respectively.
Nevertheless, our results show that the chosen partition of $C''$ is by no means arbitrary. 

We find
\begin{align*}
  C''_{I,1}=\{C_0,\dotsc,C_{29},C_{34},\dotsc,C_{66},C_{68},\dotsc,C_{82}\}, \quad C''_{I,2}=\{C_{30},C_{31},C_{32},C_{67}\}, \quad C''_{I,3}\supseteq\{C_{33}\}. 
\end{align*}
We used the Monte Carlo method to search for the set $C''_{I,3}$, and it is possible that it contains additional classes besides $C_{33}$.
However, our results show that such additional classes are very rare with respect to a measure that is uniform on the space of coefficients in linear combinations of bases vectors. 
Tables \ref{table:2222-part-1}, \ref{table:2222-part-2}, \ref{table:2222-part-3} list the classes, their independent invariants and representative elements.
In these tables, all $83$ classes appear in $27$ fundamental sets of classes related by permutations of $\{V_1,V_2,V_3,V_4\}$.
Table \ref{table:2222-permutations} lists the sets of classes and their representative elements.
\begin{table}[ht]
  \caption{\label{table:2222-part-1} The entanglement classes, their independent algebraic invariants, and their representative elements for $n=4$, $D=(2,2,2,2)$.
  Each expression $[j_1,j_2,j_3,j_4]$ stands for $u_{1,j_1}\otimes u_{2,j_2}\otimes u_{3,j_3}\otimes u_{4,j_4}$, where $\{u_{i,j}\}$ is a set of any linearly independent elements of $V_i$.
  Classes within each horizontal block are related by a permutation symmetry of $\{V_i\}_{i\in I}$.}
\footnotesize
  \begin{ruledtabular}
    \begin{tabular}{lll}
      & $\tilde{N}''(v)$ & $v$ \\ \hline
$C_0$ & $(2,2,2,2,8,8,8,8,16,16,16,16,16,16,16,16,16,16,16)$ & $0$ \\ \hline 
$C_1$ & $(1,1,1,1,4,4,4,4,10,10,10,10,10,10,8,8,8,8,11)$ & $[1,1,1,1]$ \\ \hline 
$C_2$ & $(0,0,1,1,3,3,2,2,9,7,7,7,7,10,3,3,7,7,10)$ & $[1,1,1,1]+[2,2,1,1]$ \\ 
$C_3$ & $(0,1,0,1,3,2,3,2,7,9,7,7,10,7,3,7,3,7,10)$ & $[1,1,1,1]+[2,1,2,1]$ \\ 
$C_4$ & $(0,1,1,0,2,3,3,2,7,7,9,10,7,7,3,7,7,3,10)$ & $[1,1,1,1]+[2,1,1,2]$ \\ 
$C_5$ & $(1,0,0,1,3,2,2,3,7,7,10,9,7,7,7,3,3,7,10)$ & $[1,1,1,1]+[1,2,2,1]$ \\ 
$C_6$ & $(1,0,1,0,2,3,2,3,7,10,7,7,9,7,7,3,7,3,10)$ & $[1,1,1,1]+[1,2,1,2]$ \\ 
$C_7$ & $(1,1,0,0,2,2,3,3,10,7,7,7,7,9,7,7,3,3,10)$ & $[1,1,1,1]+[1,1,2,2]$ \\ \hline 
$C_8$ & $(0,0,0,0,0,0,0,0,0,0,9,9,0,0,0,0,0,0,9)$ & $[1,1,1,1]+[1,2,2,1]+[2,1,1,2]+[2,2,2,2]$ \\ 
$C_9$ & $(0,0,0,0,0,0,0,0,0,9,0,0,9,0,0,0,0,0,9)$ & $[1,1,1,1]+[1,2,1,2]+[2,1,2,1]+[2,2,2,2]$ \\ 
$C_{10}$ & $(0,0,0,0,0,0,0,0,9,0,0,0,0,9,0,0,0,0,9)$ & $[1,1,1,1]+[1,1,2,2]+[2,2,1,1]+[2,2,2,2]$ \\ \hline 
$C_{11}$ & $(0,0,0,1,1,2,2,2,5,5,7,5,7,7,2,2,2,7,8)$ & $[1,1,1,1]+[2,1,2,1]+[2,2,1,1]$ \\ 
$C_{12}$ & $(0,0,1,0,2,1,2,2,5,7,5,7,5,7,2,2,7,2,8)$ & $[1,1,1,1]+[1,2,1,2]+[2,2,1,1]$ \\ 
$C_{13}$ & $(0,1,0,0,2,2,1,2,7,5,5,7,7,5,2,7,2,2,8)$ & $[1,1,1,1]+[1,1,2,2]+[2,1,2,1]$ \\ 
$C_{14}$ & $(1,0,0,0,2,2,2,1,7,7,7,5,5,5,7,2,2,2,8)$ & $[1,1,1,1]+[1,1,2,2]+[1,2,1,2]$ \\ \hline 
$C_{15}$ & $(0,0,0,1,0,2,2,2,4,4,7,4,7,7,2,2,2,7,7)$ & $[1,1,1,1]+[2,2,2,1]$ \\ 
$C_{16}$ & $(0,0,1,0,2,0,2,2,4,7,4,7,4,7,2,2,7,2,7)$ & $[1,1,1,1]+[2,2,1,2]$ \\ 
$C_{17}$ & $(0,1,0,0,2,2,0,2,7,4,4,7,7,4,2,7,2,2,7)$ & $[1,1,1,1]+[2,1,2,2]$ \\ 
$C_{18}$ & $(1,0,0,0,2,2,2,0,7,7,7,4,4,4,7,2,2,2,7)$ & $[1,1,1,1]+[1,2,2,2]$ \\ \hline 
$C_{19}$ & $(0,0,0,0,1,1,1,1,5,5,5,5,5,5,2,2,2,2,7)$ & $[1,1,1,1]+[2,1,1,2]+[2,1,2,1]+[2,2,1,1]$ \\ \hline 
$C_{20}$ & $(0,0,0,0,1,0,0,1,2,2,4,5,2,2,1,1,1,1,6)$ & $[1,1,1,1]+[1,2,2,1]+[2,2,1,2]$ \\ 
$C_{21}$ & $(0,0,0,0,0,1,1,0,2,2,5,4,2,2,1,1,1,1,6)$ & $[1,1,1,1]+[2,1,1,2]+[2,2,2,1]$ \\ 
$C_{22}$ & $(0,0,0,0,0,1,0,1,2,4,2,2,5,2,1,1,1,1,6)$ & $[1,1,1,1]+[1,2,1,2]+[2,2,2,1]$ \\ 
$C_{23}$ & $(0,0,0,0,1,0,1,0,2,5,2,2,4,2,1,1,1,1,6)$ & $[1,1,1,1]+[2,1,2,1]+[2,2,1,2]$ \\ 
$C_{24}$ & $(0,0,0,0,0,0,1,1,4,2,2,2,2,5,1,1,1,1,6)$ & $[1,1,1,1]+[1,1,2,2]+[2,2,2,1]$ \\ 
$C_{25}$ & $(0,0,0,0,1,1,0,0,5,2,2,2,2,4,1,1,1,1,6)$ & $[1,1,1,1]+[2,1,2,2]+[2,2,1,1]$ \\ \hline 
$C_{26}$ & $(0,0,0,0,0,0,0,0,4,4,4,4,4,4,0,0,0,0,6)$ & $[1,1,1,1]+[2,2,2,2]$ \\
    \end{tabular}
  \end{ruledtabular}
\end{table}
\begin{table}[ht]
  \caption{\label{table:2222-part-2} The entanglement classes, their independent algebraic invariants, and their representative elements for $n=4$, $D=(2,2,2,2)$.
  Each expression $[j_1,j_2,j_3,j_4]$ stands for $u_{1,j_1}\otimes u_{2,j_2}\otimes u_{3,j_3}\otimes u_{4,j_4}$, where $\{u_{i,j}\}$ is a set of any linearly independent elements of $V_i$.
  Classes within each horizontal block are related by a permutation symmetry of $\{V_i\}_{i\in I}$.}
\footnotesize
  \begin{ruledtabular}
    \begin{tabular}{lll}
      & $\tilde{N}''(v)$ & $v$ \\ \hline
$C_{27}$ & $(0,0,0,0,0,0,0,0,0,0,5,5,0,0,0,0,0,0,6)$ & $[1,1,1,1]+[1,1,2,2]+[1,2,1,2]+[2,1,2,1]+[2,2,1,1]$ \\ 
$C_{28}$ & $(0,0,0,0,0,0,0,0,0,5,0,0,5,0,0,0,0,0,6)$ & $[1,1,1,1]+[1,1,2,2]+[1,2,2,1]+[2,1,1,2]+[2,2,1,1]$ \\ 
$C_{29}$ & $(0,0,0,0,0,0,0,0,5,0,0,0,0,5,0,0,0,0,6)$ & $[1,1,1,1]+[1,2,1,2]+[1,2,2,1]+[2,1,1,2]+[2,1,2,1]$ \\ \hline
$C_{30}$ & $(0,0,0,0,0,0,0,0,0,0,1,1,0,0,0,0,0,0,6)$ & $[1,1,1,1]-[1,1,1,2]-[1,1,2,1]-[1,2,1,1]+[1,2,2,1]$ \\ & & $+[1,2,2,2]-[2,1,1,1]+[2,1,1,2]+[2,1,2,2]+[2,2,1,2]$ \\  & & $+[2,2,2,1]+[2,2,2,2]$ \\ 
$C_{31}$ & $(0,0,0,0,0,0,0,0,0,1,0,0,1,0,0,0,0,0,6)$ & $[1,1,1,1]-[1,1,1,2]-[1,1,2,1]-[1,2,1,1]+[1,2,1,2]$ \\ & & $+[1,2,2,2]-[2,1,1,1]+[2,1,2,1]+[2,1,2,2]+[2,2,1,2]$ \\ & & $+[2,2,2,1]+[2,2,2,2]$ \\ 
$C_{32}$ & $(0,0,0,0,0,0,0,0,1,0,0,0,0,1,0,0,0,0,6)$ & $[1,1,1,1]-[1,1,1,2]-[1,1,2,1]+[1,1,2,2]-[1,2,1,1]$ \\ & & $+[1,2,2,2]-[2,1,1,1]+[2,1,2,2]+[2,2,1,1]+[2,2,1,2]$ \\ & & $+[2,2,2,1]+[2,2,2,2]$ \\ \hline 
$C_{33}$ & $(0,0,0,0,0,0,0,0,0,0,0,0,0,0,0,0,0,0,6)$ & $[1,1,1,1]+c[1,1,2,2]-(1+c)[1,2,1,2]-(1+c)[2,1,2,1]$ \\ & & $+c[2,2,1,1]+[2,2,2,2]$, $c\in F$, $c\not\in\{-2,-1,0,1\}$ \\ \hline
$C_{34}$ & $(0,0,0,0,0,0,0,1,2,2,2,2,2,2,0,1,1,1,5)$ & $[1,1,1,1]+[1,2,2,1]+[2,1,1,1]+[2,2,1,2]$ \\ 
$C_{35}$ & $(0,0,0,0,0,0,1,0,2,2,2,2,2,2,1,0,1,1,5)$ & $[1,1,1,1]+[1,1,2,2]+[1,2,1,1]+[2,2,2,1]$ \\
$C_{36}$ & $(0,0,0,0,0,1,0,0,2,2,2,2,2,2,1,1,0,1,5)$ & $[1,1,1,1]+[1,1,2,1]+[1,2,2,2]+[2,1,1,2]$ \\
$C_{37}$ & $(0,0,0,0,1,0,0,0,2,2,2,2,2,2,1,1,1,0,5)$ & $[1,1,1,1]+[1,1,1,2]+[2,1,2,2]+[2,2,1,1]$ \\ \hline 
$C_{38}$ & $(0,0,0,0,0,0,0,0,2,2,4,4,2,2,0,0,0,0,5)$ & $[1,1,1,1]+[1,2,2,1]+[2,1,1,2]$ \\ 
$C_{39}$ & $(0,0,0,0,0,0,0,0,2,4,2,2,4,2,0,0,0,0,5)$ & $[1,1,1,1]+[1,2,1,2]+[2,1,2,1]$ \\ 
$C_{40}$ & $(0,0,0,0,0,0,0,0,4,2,2,2,2,4,0,0,0,0,5)$ & $[1,1,1,1]+[1,1,2,2]+[2,2,1,1]$ \\ \hline
$C_{41}$ & $(0,0,0,0,0,0,0,0,0,0,4,5,0,0,0,0,0,0,5)$ & $[1,1,1,1]+[1,1,2,1]+[1,2,1,1]+[2,1,2,2]+[2,2,1,2]$ \\ 
$C_{42}$ & $(0,0,0,0,0,0,0,0,0,0,5,4,0,0,0,0,0,0,5)$ & $[1,1,1,1]+[1,1,1,2]+[1,2,2,2]+[2,1,1,1]+[2,2,2,1]$ \\ 
$C_{43}$ & $(0,0,0,0,0,0,0,0,0,4,0,0,5,0,0,0,0,0,5)$ & $[1,1,1,1]+[1,1,1,2]+[1,2,1,1]+[2,1,2,2]+[2,2,2,1]$ \\ 
$C_{44}$ & $(0,0,0,0,0,0,0,0,0,5,0,0,4,0,0,0,0,0,5)$ & $[1,1,1,1]+[1,1,2,1]+[1,2,2,2]+[2,1,1,1]+[2,2,1,2]$ \\ 
$C_{45}$ & $(0,0,0,0,0,0,0,0,4,0,0,0,0,5,0,0,0,0,5)$ & $[1,1,1,1]+[1,1,1,2]+[1,1,2,1]+[2,2,1,2]+[2,2,2,1]$ \\ 
$C_{46}$ & $(0,0,0,0,0,0,0,0,5,0,0,0,0,4,0,0,0,0,5)$ & $[1,1,1,1]+[1,2,1,1]+[1,2,2,2]+[2,1,1,1]+[2,1,2,2]$ \\ \hline 
$C_{47}$ & $(0,0,0,0,0,0,0,0,0,0,4,4,0,0,0,0,0,0,5)$ & $[1,1,1,1]+[1,2,2,2]+[2,1,1,2]+[2,2,2,1]$ \\ 
$C_{48}$ & $(0,0,0,0,0,0,0,0,0,4,0,0,4,0,0,0,0,0,5)$ & $[1,1,1,1]+[1,2,2,2]+[2,1,2,1]+[2,2,1,2]$ \\ 
$C_{49}$ & $(0,0,0,0,0,0,0,0,4,0,0,0,0,4,0,0,0,0,5)$ & $[1,1,1,1]+[1,2,2,2]+[2,1,2,2]+[2,2,1,1]$ \\ \hline 
$C_{50}$ & $(0,0,0,0,0,0,0,0,1,1,2,1,2,2,0,0,0,1,4)$ & $[1,1,1,1]+[1,2,2,2]+[2,1,2,1]+[2,2,1,1]$ \\ 
$C_{51}$ & $(0,0,0,0,0,0,0,0,1,2,1,2,1,2,0,0,1,0,4)$ & $[1,1,1,1]+[1,2,1,2]+[2,1,2,2]+[2,2,1,1]$ \\
$C_{52}$ & $(0,0,0,0,0,0,0,0,2,1,1,2,2,1,0,1,0,0,4)$ & $[1,1,1,1]+[1,1,2,2]+[2,1,2,1]+[2,2,1,2]$ \\
$C_{53}$ & $(0,0,0,0,0,0,0,0,2,2,2,1,1,1,1,0,0,0,4)$ & $[1,1,1,1]+[1,1,2,2]+[1,2,1,2]+[2,2,2,1]$ \\ \hline 
$C_{54}$ & $(0,0,0,0,0,0,0,0,1,2,2,2,2,2,0,0,0,0,4)$ & $[1,1,1,1]+[1,2,1,2]+[2,1,2,1]+[2,2,1,1]$ \\ 
$C_{55}$ & $(0,0,0,0,0,0,0,0,2,1,2,2,2,2,0,0,0,0,4)$ & $[1,1,1,1]+[1,1,2,2]+[2,1,2,1]+[2,2,1,1]$ \\ 
$C_{56}$ & $(0,0,0,0,0,0,0,0,2,2,1,2,2,2,0,0,0,0,4)$ & $[1,1,1,1]+[1,1,2,2]+[2,1,1,2]+[2,2,1,1]$ \\ 
$C_{57}$ & $(0,0,0,0,0,0,0,0,2,2,2,1,2,2,0,0,0,0,4)$ & $[1,1,1,1]+[1,1,2,2]+[1,2,2,1]+[2,2,1,1]$ \\ 
$C_{58}$ & $(0,0,0,0,0,0,0,0,2,2,2,2,1,2,0,0,0,0,4)$ & $[1,1,1,1]+[1,1,2,2]+[1,2,1,2]+[2,2,1,1]$ \\ 
$C_{59}$ & $(0,0,0,0,0,0,0,0,2,2,2,2,2,1,0,0,0,0,4)$ & $[1,1,1,1]+[1,1,2,2]+[1,2,1,2]+[2,1,2,1]$ \\
    \end{tabular}
  \end{ruledtabular}
\end{table}
\begin{table}[ht]
  \caption{\label{table:2222-part-3} The entanglement classes, their independent algebraic invariants, and their representative elements for $n=4$, $D=(2,2,2,2)$.
  Each expression $[j_1,j_2,j_3,j_4]$ stands for $u_{1,j_1}\otimes u_{2,j_2}\otimes u_{3,j_3}\otimes u_{4,j_4}$, where $\{u_{i,j}\}$ is a set of any linearly independent elements of $V_i$.
  Classes within each horizontal block are related by a permutation symmetry of $\{V_i\}_{i\in I}$.}
\footnotesize
  \begin{ruledtabular}
    \begin{tabular}{lll}
      & $\tilde{N}''(v)$ & $v$ \\ \hline
$C_{60}$ & $(0,0,0,0,0,0,0,0,1,1,1,1,1,1,0,0,0,0,4)$ & $[1,1,1,1]+[1,2,1,2]+[1,2,2,1]+[2,1,1,2]+[2,1,2,1]$ \\ & & $+[2,2,2,2]$ \\ \hline 
$C_{61}$ & $(0,0,0,0,0,0,0,0,0,0,2,2,0,0,0,0,0,0,4)$ & $[1,1,1,1]+[1,1,2,2]+[1,2,1,1]+[2,1,1,1]+[2,1,2,1]$ \\ & & $+[2,2,1,2]$ \\
$C_{62}$ & $(0,0,0,0,0,0,0,0,0,2,0,0,2,0,0,0,0,0,4)$ & $[1,1,1,1]+[1,1,1,2]+[1,2,2,1]+[2,1,1,1]+[2,1,2,2]$ \\ & & $+[2,2,1,1]$ \\
$C_{63}$ & $(0,0,0,0,0,0,0,0,2,0,0,0,0,2,0,0,0,0,4)$ & $[1,1,1,1]+[1,1,2,1]+[1,2,1,2]+[2,1,1,1]+[2,1,1,2]$ \\ & & $+[2,2,2,1]$ \\ \hline
$C_{64}$ & $(0,0,0,0,0,0,0,0,0,0,1,1,0,0,0,0,0,0,4)$ & $[1,1,1,1]+[1,1,2,2]+[1,2,1,1]+[1,2,1,2]+[1,2,2,1]$ \\ & & $+[2,1,1,1]+[2,1,1,2]+[2,1,2,2]+[2,2,2,1]+[2,2,2,2]$ \\
$C_{65}$ & $(0,0,0,0,0,0,0,0,0,1,0,0,1,0,0,0,0,0,4)$ & $[1,1,1,1]+[1,1,1,2]+[1,1,2,2]+[1,2,1,2]+[1,2,2,1]$ \\ & & $+[2,1,1,1]+[2,1,2,1]+[2,2,1,2]+[2,2,2,1]+[2,2,2,2]$ \\
$C_{66}$ & $(0,0,0,0,0,0,0,0,1,0,0,0,0,1,0,0,0,0,4)$ & $[1,1,1,1]+[1,1,2,1]+[1,1,2,2]+[1,2,1,2]+[1,2,2,1]$ \\ & & $+[2,1,1,1]+[2,1,2,2]+[2,2,1,1]+[2,2,1,2]+[2,2,2,2]$ \\ \hline
$C_{67}$ & $(0,0,0,0,0,0,0,0,0,0,0,0,0,0,0,0,0,0,4)$ & $[1,1,1,1]+[1,1,1,2]+[1,1,2,1]-[1,2,1,1]+[1,2,1,2]$ \\ & & $-[2,1,1,1]+[2,1,2,1]+[2,2,1,1]+[2,2,2,2]$ \\ \hline 
$C_{68}$ & $(0,0,0,0,0,0,0,0,1,1,2,1,2,2,0,0,0,0,3)$ & $[1,1,1,1]+[1,1,2,1]+[1,2,1,1]+[2,1,1,1]+[2,2,2,2]$ \\ 
$C_{69}$ & $(0,0,0,0,0,0,0,0,1,2,1,2,1,2,0,0,0,0,3)$ & $[1,1,1,1]+[1,1,1,2]+[1,2,1,1]+[2,1,1,1]+[2,2,2,2]$ \\ 
$C_{70}$ & $(0,0,0,0,0,0,0,0,2,1,1,2,2,1,0,0,0,0,3)$ & $[1,1,1,1]+[1,1,1,2]+[1,1,2,1]+[2,1,1,1]+[2,2,2,2]$ \\ 
$C_{71}$ & $(0,0,0,0,0,0,0,0,2,2,2,1,1,1,0,0,0,0,3)$ & $[1,1,1,1]+[1,1,1,2]+[1,1,2,1]+[1,2,1,1]+[2,2,2,2]$ \\ \hline 
$C_{72}$ & $(0,0,0,0,0,0,0,0,1,1,1,1,1,1,0,0,0,0,3)$ & $[1,1,1,1]+[1,2,2,2]+[2,1,1,2]+[2,1,2,1]+[2,2,1,1]$ \\ \hline
$C_{73}$ & $(0,0,0,0,0,0,0,0,0,0,1,2,0,0,0,0,0,0,3)$ & $[1,1,1,1]+[1,1,2,2]+[1,2,1,1]+[2,1,2,1]+[2,2,1,2]$ \\ 
$C_{74}$ & $(0,0,0,0,0,0,0,0,0,0,2,1,0,0,0,0,0,0,3)$ & $[1,1,1,1]+[1,1,2,2]+[1,2,1,2]+[2,1,1,1]+[2,2,2,1]$ \\ 
$C_{75}$ & $(0,0,0,0,0,0,0,0,0,1,0,0,2,0,0,0,0,0,3)$ & $[1,1,1,1]+[1,1,2,2]+[1,2,1,1]+[2,1,1,2]+[2,2,2,1]$ \\ 
$C_{76}$ & $(0,0,0,0,0,0,0,0,0,2,0,0,1,0,0,0,0,0,3)$ & $[1,1,1,1]+[1,1,2,2]+[1,2,2,1]+[2,1,1,1]+[2,2,1,2]$ \\ 
$C_{77}$ & $(0,0,0,0,0,0,0,0,1,0,0,0,0,2,0,0,0,0,3)$ & $[1,1,1,1]+[1,1,2,1]+[1,2,1,2]+[2,1,1,2]+[2,2,2,1]$ \\ 
$C_{78}$ & $(0,0,0,0,0,0,0,0,2,0,0,0,0,1,0,0,0,0,3)$ & $[1,1,1,1]+[1,2,1,2]+[1,2,2,1]+[2,1,1,1]+[2,1,2,2]$ \\ \hline 
$C_{79}$ & $(0,0,0,0,0,0,0,0,0,0,1,1,0,0,0,0,0,0,3)$ & $[1,1,1,1]+[1,1,1,2]+[1,2,2,1]+[2,1,1,1]+[2,1,2,2]$ \\ & & $+[2,2,1,2]$ \\
$C_{80}$ & $(0,0,0,0,0,0,0,0,0,1,0,0,1,0,0,0,0,0,3)$ & $[1,1,1,1]+[1,1,2,1]+[1,2,1,2]+[2,1,1,1]+[2,1,2,2]$ \\ & & $+[2,2,2,1]$ \\
$C_{81}$ & $(0,0,0,0,0,0,0,0,1,0,0,0,0,1,0,0,0,0,3)$ & $[1,1,1,1]+[1,1,2,2]+[1,2,1,1]+[2,1,1,1]+[2,2,1,2]$ \\ & & $+[2,2,2,1]$ \\ \hline 
$C_{82}$ & $(0,0,0,0,0,0,0,0,0,0,0,0,0,0,0,0,0,0,3)$ & $[1,1,1,1]+[1,1,2,2]+[1,2,1,1]+[1,2,1,2]+[2,1,1,1]$ \\ & & $+[2,1,2,1]+[2,2,2,2]$ \\
    \end{tabular}
  \end{ruledtabular}
\end{table}
\begin{table}[ht]
  \caption{\label{table:2222-permutations} Representative elements for the sets of equivalence classes for $n=4$, $D=(2,2,2,2)$ induced by the permutation symmetry of the spaces in $\{V_1,\dotsc,V_4\}$.
  A representative element $v$ is given by $v=Av_1$, where $A\colon V\to V$ is a certain linear operator and $v_1\in V$ is a fixed vector. 
  (Without loss of generality and for comparison with other tables, we choose $v_1=[1,1,1,1]$.)
  The operator $a_i$ is defined by $a_i[\dotsc,1,\dotsc]=[\dotsc,2,\dotsc]$ and $a_i[\dotsc,2,\dotsc]=[\dotsc,1,\dotsc]$, where only the $i$th index changes.
  To obtain all classes in each group, all possible choices of the indices $\{i,j,k,l\}=\{1,2,3,4\}$ should be considered.
  This results in $27$ fundamental sets of $83$ classes.}
\footnotesize
  \begin{ruledtabular}
    \begin{tabular}{ll}
      & $A$ \\ \hline
      $C_{0}$                    & $0$ \\
      $C_{1}$                    & $1$ \\
      $\{C_{2},\ldots,C_{7}\}$   & $1+a_i a_j$ \\
      $\{C_{8},C_9,C_{10}\}$     & $(1+a_i a_j)(1+a_k a_l)$  \\
      $\{C_{11},\ldots,C_{14}\}$ & $1+a_i(a_j+a_k)$ \\
      $\{C_{15},\ldots,C_{18}\}$ & $1+a_i a_j a_k$ \\
      $C_{19}$                   & $1+a_i(a_j+a_k+a_l)$ \\
      $\{C_{20},\ldots,C_{25}\}$ & $1+a_i(a_j+a_k a_l)$ \\
      $C_{26}$                   & $1+a_i a_j a_k a_l$ \\
      $\{C_{27},C_{28},C_{29}\}$ & $1+(a_i+a_j)(a_k+a_l)$ \\
      $\{C_{30},C_{31},C_{32}\}$ & $1-(a_i+a_j+a_k+a_l)+a_j a_k a_l+a_i a_k a_l+a_i a_j a_l+a_i a_j a_k+a_i a_j+a_k a_l+a_i a_j a_k a_l$ \\
      $C_{33}$                   & $1+c(a_i a_j+a_k a_l)-(1+c)(a_i a_k+a_j a_l)+a_ i a_j a_k a_l$, $c\in F$, $c\not\in\{-2,-1,0,1\}$ \\
      $\{C_{34},\ldots,C_{37}\}$ & $1+a_i+a_j a_k+a_i a_k a_l$ \\
      $\{C_{38},C_{39},C_{40}\}$ & $1+a_i a_j+a_k a_l$ \\
      $\{C_{41},\ldots,C_{46}\}$ & $1+(a_i+a_j)(1+a_k a_l)$ \\
      $\{C_{47},C_{48},C_{49}\}$ & $1+a_i a_j+(a_i+a_j)a_k a_l$ \\
      $\{C_{50},\ldots,C_{53}\}$ & $1+a_i(a_j+a_k)+a_j a_k a_l$ \\
      $\{C_{54},\ldots,C_{59}\}$ & $1+a_i a_j+a_k a_l+a_i a_k$ \\
      $C_{60}$                   & $1+(a_i+a_j)(a_k+a_l)+a_i a_j a_k a_l$ \\
      $\{C_{61},C_{62},C_{63}\}$ & $1+a_i+a_j+a_k a_l+a_i a_k+a_i a_j a_l$ \\
      $\{C_{64},C_{65},C_{66}\}$ & $1+a_i+a_j+a_k a_l+a_i a_l+a_j a_k+a_j a_l+a_i a_k a_l+a_i a_j a_k+a_i a_j a_k a_l$ \\
      $C_{67}$                   & $1-a_i-a_j+a_k+a_l+a_i a_j+a_i a_k+a_j a_l +a_i a_j a_k a_l$ \\
      $\{C_{68},\ldots,C_{71}\}$ & $1+a_i+a_j+a_k+a_i a_j a_k a_l$ \\
      $C_{72}$                   & $1+a_i(a_j+a_k+a_l)+a_j a_k a_l$ \\
      $\{C_{73},\ldots,C_{78}\}$ & $1+a_i(1+a_k a_l)+a_j(a_k+a_l)$ \\
      $\{C_{79},C_{80},C_{81}\}$ & $1+a_i+a_j+a_k a_l+a_i a_j(a_k+a_l)$ \\
      $C_{82}$                   & $1+a_i+a_j+a_k a_l+a_i a_k+a_j a_l +a_i a_j a_k a_l$ \\
    \end{tabular}
  \end{ruledtabular}
\end{table}

\section{Comparison with classical invariants}

The central distinction between the classification presented in this work and classifications found in the literature is in the type of invariants used.
We rely on discrete invariants, while most other methods use continuous invariants.
Broadly speaking, the relation between the two types of invariants is such that different values of the discrete invariants correspond to certain continuous invariants being zero or nonzero.
We do not attempt here the complete comparison between classifications based on the two types of invariants and present results only for the methods reviewed in \cite{Borsten:2008wd} for $3$ qubits and developed in \cite{Luque1} for $4$ qubits.

\subsection{$\bm{n=3}$}

For an arbitrary state of $3$ qubits
\begin{align*}
  v=\sum_{j_1=1}^2\sum_{j_2=1}^2\sum_{j_3=1}^2 v_{j_1,j_2,j_3}e_{1,j_1}\otimes e_{2,j_2}\otimes e_{3,j_3},
  \label{}
\end{align*}
the classical invariants (see e.g.~\cite{Borsten:2008wd}) are 
\begin{align*}
  h_1&=v_{1,1,1}v_{1,2,2}-v_{1,1,2}v_{1,2,1}+v_{2,1,1}v_{2,2,2}-v_{2,1,2}v_{2,2,1}, \\
  h_2&=v_{1,1,1}v_{2,1,2}-v_{1,1,2}v_{2,1,1}+v_{1,2,1}v_{2,2,2}-v_{1,2,2}v_{2,2,1}, \\
  h_3&=v_{1,1,1}v_{2,2,1}-v_{1,2,1}v_{2,1,1}+v_{1,1,2}v_{2,2,2}-v_{1,2,2}v_{2,1,2}, \\
  h_4&=v_{1,1,1}^2 v_{2,2,2}^2+v_{1,1,2}^2 v_{2,2,1}^2+v_{1,2,1}^2 v_{2,1,2}^2+v_{2,1,1}^2 v_{1,2,2}^2 \\ &-2(v_{1,1,1}v_{1,1,2}v_{2,2,1}v_{2,2,2}+v_{1,1,1}v_{1,1,2}v_{2,2,1}v_{2,2,2}+v_{1,1,1}v_{1,1,2}v_{2,2,1}v_{2,2,2} \\ &+v_{1,1,1}v_{1,1,2}v_{2,2,1}v_{2,2,2}+v_{1,1,1,1}v_{1,1,2}v_{2,2,1}v_{2,2,2}+v_{1,1,1}v_{1,1,2}v_{2,2,1}v_{2,2,2}) \\ &+4(v_{1,1,1}v_{1,2,2}v_{2,1,2}v_{2,2,1}+v_{1,1,2}v_{1,2,1}v_{2,1,1}v_{2,2,2}).
\end{align*}
Table \ref{table:continuous-invariants-222} lists zero and nonzero values of $h_1,\ldots,h_4$ for the classes $C_0,\ldots,C_6$.
Ignoring the trivial difference between $C_0$ and $C_1$, we conclude that zero values of the continuous invariants $h_1,\ldots,h_4$ distinguish all the classes for $3$ qubits found using the discrete invariants.
Similar comparisons for $n=3$, $D=(d_1,d_2,d_3)$ can be easily done.

\begin{table}[ht]
  \caption{\label{table:continuous-invariants-222} The continuous invariants $h_1,\ldots,h_4$ for the sets of equivalence classes based on the discrete invariants for $n=3$, $D=(2,2,2)$.}
\footnotesize
  \begin{ruledtabular}
    \begin{tabular}{lcccc}
      & $h_1$ & $h_2$ & $h_3$ & $h_4$ \\ \hline
      $C_{0}$ & $0$ & $0$ & $0$ & $0$ \\ \hline
      $C_{1}$ & $0$ & $0$ & $0$ & $0$ \\ \hline
      $C_{2}$ & $0$ & $0$ & $\not=0$ & $0$ \\
      $C_{3}$ & $0$ & $\not=0$ & $0$ & $0$ \\
      $C_{4}$ & $\not=0$ & $0$ & $0$ & $0$ \\ \hline
      $C_{5}$ & $\not=0$ & $\not=0$ & $\not=0$ & $0$ \\ \hline
      $C_{6}$ & $0$ & $0$ & $0$ & $\not=0$ \\
    \end{tabular}
  \end{ruledtabular}
\end{table}
\subsection{$\bm{n=4}$}

For an arbitrary state of $4$ qubits
\begin{align*}
  v=\sum_{j_1=1}^2\cdots\sum_{j_4=1}^2 v_{j_1,\ldots,j_4}e_{1,j_1}\otimes\cdots\otimes e_{4,j_4},
  \label{}
\end{align*}
the Hilbert series lead \cite{Luque1} to the classical invariants
\begin{align*}
  h_1&=v_{1,1,1,1}v_{2,2,2,2}-v_{1,1,1,2}v_{2,2,2,1}-v_{1,1,2,1}v_{2,2,1,2}+v_{1,1,2,2}v_{2,2,1,1}\nn\\
  &-v_{1,2,1,1}v_{2,1,2,2}+v_{1,2,1,2}v_{2,1,2,1}+v_{1,2,2,1}v_{2,1,1,2}-v_{1,2,2,2}v_{2,1,1,1},\\
  h_2&=
  \begin{vmatrix}
    v_{1,1,1,1} & v_{1,2,1,1} & v_{2,1,1,1} & v_{2,2,1,1}\\
    v_{1,1,1,2} & v_{1,2,1,2} & v_{2,1,1,2} & v_{2,2,1,2}\\
    v_{1,1,2,1} & v_{1,2,2,1} & v_{2,1,2,1} & v_{2,2,2,1}\\
    v_{1,1,2,2} & v_{1,2,2,2} & v_{2,1,2,2} & v_{2,2,2,2}\\
  \end{vmatrix}
  ,\\
  h_3&=
  \begin{vmatrix}
    v_{1,1,1,1} & v_{2,1,1,1} & v_{1,1,2,1} & v_{2,1,2,1}\\
    v_{1,1,1,2} & v_{2,1,1,2} & v_{1,1,2,2} & v_{2,1,2,2}\\
    v_{1,2,1,1} & v_{2,2,1,1} & v_{1,2,2,1} & v_{2,2,2,1}\\
    v_{1,2,1,2} & v_{2,2,1,2} & v_{1,2,2,2} & v_{2,2,2,2}\\
  \end{vmatrix}
  ,\\
  h_4&=
  \begin{vmatrix}
    v_{1,1,1,1} & v_{1,1,1,2} & v_{2,1,1,1} & v_{2,1,1,2}\\
    v_{1,1,2,1} & v_{1,1,2,2} & v_{2,1,2,1} & v_{2,1,2,2}\\
    v_{1,2,1,1} & v_{1,2,1,2} & v_{2,2,1,1} & v_{2,2,1,2}\\
    v_{1,2,2,1} & v_{1,2,2,2} & v_{2,2,2,1} & v_{2,2,2,2}\\
  \end{vmatrix}
  ,\\
  h_5&=\det(h_{5,i,j})_{1\le i\le 3;1\le j\le 3},\\
  h_{5,1,1}&=-v_{1,1,1,2}v_{1,1,2,1}+v_{1,1,1,1}v_{1,1,2,2},\\
  h_{5,1,2}&= v_{1,1,2,2}v_{1,2,1,1}-v_{1,1,2,1}v_{1,2,1,2}-v_{1,1,1,2}v_{1,2,2,1}+v_{1,1,1,1}v_{1,2,2,2},\\
  h_{5,1,3}&=-v_{1,2,1,2}v_{1,2,2,1}+v_{1,2,1,1}v_{1,2,2,2},\\
  h_{5,2,1}&= v_{1,1,2,2}v_{2,1,1,1}-v_{1,1,2,1}v_{2,1,1,2}-v_{1,1,1,2}v_{2,1,2,1}+v_{1,1,1,1}v_{2,1,2,2},\\
  h_{5,2,2}&= v_{1,2,2,2}v_{2,1,1,1}-v_{1,2,2,1}v_{2,1,1,2}-v_{1,2,1,2}v_{2,1,2,1}+v_{1,2,1,1}v_{2,1,2,2} \\ 
  &+v_{1,1,2,2}v_{2,2,1,1}-v_{1,1,2,1}v_{2,2,1,2}-v_{1,1,1,2}v_{2,2,2,1}+v_{1,1,1,1}v_{2,2,2,2},\\
  h_{5,2,3}&= v_{1,2,2,2}v_{2,2,1,1}-v_{1,2,2,1}v_{2,2,1,2}-v_{1,2,1,2}v_{2,2,2,1}+v_{1,2,1,1}v_{2,2,2,2},\\
  h_{5,3,1}&=-v_{2,1,1,2}v_{2,1,2,1}+v_{2,1,1,1}v_{2,1,2,2},\\
  h_{5,3,2}&= v_{2,1,2,2}v_{2,2,1,1}-v_{2,1,2,1}v_{2,2,1,2}-v_{2,1,1,2}v_{2,2,2,1}+v_{2,1,1,1}v_{2,2,2,2},\\
  h_{5,3,3}&=-v_{2,2,1,2}v_{2,2,2,1}+v_{2,2,1,1}v_{2,2,2,2},\\
  h_6&=\det(h_{6,i,j})_{1\le i\le 3;1\le j\le 3},\\
  h_{6,1,1}&=-v_{1,1,1,2} v_{1,2,1,1}+v_{1,1,1,1}v_{1,2,1,2},\\
  h_{6,1,2}&=-v_{1,1,2,2}v_{1,2,1,1}+v_{1,1,2,1}v_{1,2,1,2}-v_{1,1,1,2}v_{1,2,2,1}+v_{1,1,1,1}v_{1,2,2,2},\\
  h_{6,1,3}&=-v_{1,1,2,2}v_{1,2,2,1}+v_{1,1,2,1}v_{1,2,2,2},\\
  h_{6,2,1}&= v_{1,2,1,2}v_{2,1,1,1}-v_{1,2,1,1}v_{2,1,1,2}-v_{1,1,1,2}v_{2,2,1,1}+v_{1,1,1,1}v_{2,2,1,2},\\
  h_{6,2,2}&= v_{1,2,2,2}v_{2,1,1,1}-v_{1,2,2,1}v_{2,1,1,2}+v_{1,2,1,2}v_{2,1,2,1}-v_{1,2,1,1}v_{2,1,2,2} \\
  &-v_{1,1,2,2}v_{2,2,1,1}+v_{1,1,2,1}v_{2,2,1,2}-v_{1,1,1,2}v_{2,2,2,1}+v_{1,1,1,1}v_{2,2,2,2},\\
  h_{6,2,3}&= v_{1,2,2,2}v_{2,1,2,1}-v_{1,2,2,1}v_{2,1,2,2}-v_{1,1,2,2}v_{2,2,2,1}+v_{1,1,2,1}v_{2,2,2,2},\\
  h_{6,3,1}&=-v_{2,1,1,2}v_{2,2,1,1}+v_{2,1,1,1}v_{2,2,1,2},\\
  h_{6,3,2}&=-v_{2,1,2,2}v_{2,2,1,1}+v_{2,1,2,1}v_{2,2,1,2}-v_{2,1,1,2}v_{2,2,2,1}+v_{2,1,1,1}v_{2,2,2,2},\\
  h_{6,3,3}&=-v_{2,1,2,2}v_{2,2,2,1}+v_{2,1,2,1}v_{2,2,2,2},\\
  h_7&=\det(h_{7,i,j})_{1\le i\le 3;1\le j\le 3},\\
  h_{7,1,1}&=-v_{1,1,2,1}v_{1,2,1,1}+v_{1,1,1,1}v_{1,2,2,1},\\
  h_{7,1,2}&=-v_{1,1,2,2}v_{1,2,1,1}-v_{1,1,2,1}v_{1,2,1,2}+ v_{1,1,1,2}v_{1,2,2,1}+v_{1,1,1,1}v_{1,2,2,2},\\
  h_{7,1,3}&=-v_{1,1,2,2}v_{1,2,1,2}+v_{1,1,1,2}v_{1,2,2,2},\\
  h_{7,2,1}&= v_{1,2,2,1}v_{2,1,1,1}-v_{1,2,1,1}v_{2,1,2,1}-v_{1,1,2,1}v_{2,2,1,1}+v_{1,1,1,1}v_{2,2,2,1},\\
  h_{7,2,2}&= v_{1,2,2,2}v_{2,1,1,1}+v_{1,2,2,1}v_{2,1,1,2}-v_{1,2,1,2}v_{2,1,2,1}-v_{1,2,1,1}v_{2,1,2,2} \\
  &-v_{1,1,2,2}v_{2,2,1,1}-v_{1,1,2,1}v_{2,2,1,2}+v_{1,1,1,2}v_{2,2,2,1}+v_{1,1,1,1}v_{2,2,2,2},\\
  h_{7,2,3}&= v_{1,2,2,2}v_{2,1,1,2}-v_{1,2,1,2}v_{2,1,2,2}-v_{1,1,2,2}v_{2,2,1,2}+v_{1,1,1,2}v_{2,2,2,2},\\
  h_{7,3,1}&=-v_{2,1,2,1}v_{2,2,1,1}+v_{2,1,1,1}v_{2,2,2,1},\\
  h_{7,3,2}&=-v_{2,1,2,2}v_{2,2,1,1}-v_{2,1,2,1}v_{2,2,1,2}+ v_{2,1,1,2}v_{2,2,2,1}+v_{2,1,1,1}v_{2,2,2,2},\\
  h_{7,3,3}&=-v_{2,1,2,2}v_{2,2,1,2}+v_{2,1,1,2}v_{2,2,2,2}.\\
  \label{}
\end{align*}
The invariants satisfy the relations
\begin{align*}
  h_2+h_3+h_4&=0, \\
  h_1 h_2-h_6+h_7&=0, \\  
  h_1 h_3-h_7+h_5&=0, \\  
  h_1 h_4-h_5+h_6&=0,
\end{align*}
which imply that only two invariants among $h_2,h_3,h_4$ are independent and only one invariant among $h_5,h_6,h_7$ is independent. 
Nevertheless, in the results below we use all the invariants $h_1,\ldots,h_7$ for the sake of symmetry.

Tables \ref{table:continuous-invariants-2222-1}, \ref{table:continuous-invariants-2222-2}, \ref{table:continuous-invariants-2222-3} list zero and nonzero values of $h_1,\ldots,h_7$ for the classes $C_0,\ldots,C_{82}$.
Table \ref{table:continuous-invariants-2222-4} shows that with only $21$ possibilities for their independent values, the zero values of invariants $h_1,\ldots,h_7$ cannot distinguish all the classes found using the discrete invariants.
Of course, since $h_1,\ldots,h_7$ is a complete list of invariants for $4$ qubits, all their possible values completely characterize the entangled states (and with a greater refinement than our invariants, as explained earlier).
The difficulty of using the continuous invariants is of course in finding all their allowed values.

Note that some of our classes split with respect to the values of the classical invariants, but it should be remembered that there are relations that could allow individual classical invariants to be less constrained than their irreducible set.

We also note that the families of entangled $4$ qubits found in \cite{Verstraete} are related to our entanglement classes as follows:
\begin{align*}
  G_{abcd}\in C_{82}, \quad L_{abc_2}\in C_{82}, \quad L_{a_2 b_2}\in\{C_{73},\ldots,C_{78}\}, \quad 
  L_{ab_3}\in C_{67}, \\ L_{a_4}\in\{C_{73},\ldots,C_{78}\}, \quad L_{a_2 0_{3\oplus\overline{1}}}\in \{C_{68},\ldots,C_{71}\}, \quad L_{0_{5\oplus\overline{3}}}\in\{C_{34},\ldots,C_{37}\}, \\ L_{0_{7\oplus\overline{1}}}\in\{C_{50},\ldots,C_{53}\}, \quad L_{0_{3\oplus\overline{1}}0_{3\oplus\overline{1}}}\in\{C_{15},\ldots,C_{18}\}.
\end{align*}
With the correction suggested in \cite{Chterental}, we find the same result as above except now $L_{ab_3}\in C_{82}$. 
This leaves many classes found here unaccounted for in the method of \cite{Verstraete}.

\begin{table}[ht]
  \caption{\label{table:continuous-invariants-2222-1} The continuous invariants $h_1,\ldots,h_7$ for the sets of equivalence classes based on the discrete invariants for $n=4$, $D=(2,2,2,2)$.}
\footnotesize
  \begin{ruledtabular}
    \begin{tabular}{lccccccc}
      & $h_1$ & $h_2$ & $h_3$ & $h_4$ & $h_5$ & $h_6$ & $h_7$ \\ \hline
      $C_{0}$ & $0$ & $0$ & $0$ & $0$ & $0$ & $0$ & $0$ \\ \hline
      $C_{1}$ & $0$ & $0$ & $0$ & $0$ & $0$ & $0$ & $0$ \\ \hline
      $C_{2}$ & $0$ & $0$ & $0$ & $0$ & $0$ & $0$ & $0$ \\
      $C_{3}$ & $0$ & $0$ & $0$ & $0$ & $0$ & $0$ & $0$ \\
      $C_{4}$ & $0$ & $0$ & $0$ & $0$ & $0$ & $0$ & $0$ \\
      $C_{5}$ & $0$ & $0$ & $0$ & $0$ & $0$ & $0$ & $0$ \\
      $C_{6}$ & $0$ & $0$ & $0$ & $0$ & $0$ & $0$ & $0$ \\
      $C_{7}$ & $0$ & $0$ & $0$ & $0$ & $0$ & $0$ & $0$ \\ \hline
      $C_{8}$ & $\not=0$ & $\not=0$ & $\not=0$ & $0$ & $0$ & $0$ & $\not=0$ \\
      $C_{9}$ & $\not=0$ & $\not=0$ & $0$ & $\not=0$ & $0$ & $\not=0$ & $0$ \\
      $C_{10}$ & $\not=0$ & $0$ & $\not=0$ & $\not=0$ & $\not=0$ & $0$ & $0$ \\ \hline
      $C_{11}$ & $0$ & $0$ & $0$ & $0$ & $0$ & $0$ & $0$ \\
      $C_{12}$ & $0$ & $0$ & $0$ & $0$ & $0$ & $0$ & $0$ \\
      $C_{13}$ & $0$ & $0$ & $0$ & $0$ & $0$ & $0$ & $0$ \\
      $C_{14}$ & $0$ & $0$ & $0$ & $0$ & $0$ & $0$ & $0$ \\ \hline
      $C_{15}$ & $0$ & $0$ & $0$ & $0$ & $0$ & $0$ & $0$ \\
      $C_{16}$ & $0$ & $0$ & $0$ & $0$ & $0$ & $0$ & $0$ \\
      $C_{17}$ & $0$ & $0$ & $0$ & $0$ & $0$ & $0$ & $0$ \\
      $C_{18}$ & $0$ & $0$ & $0$ & $0$ & $0$ & $0$ & $0$ \\ \hline
      $C_{19}$ & $0$ & $0$ & $0$ & $0$ & $0$ & $0$ & $0$ \\ \hline
      $C_{20}$ & $0$ & $0$ & $0$ & $0$ & $0$ & $0$ & $0$ \\
      $C_{21}$ & $0$ & $0$ & $0$ & $0$ & $0$ & $0$ & $0$ \\
      $C_{22}$ & $0$ & $0$ & $0$ & $0$ & $0$ & $0$ & $0$ \\
      $C_{23}$ & $0$ & $0$ & $0$ & $0$ & $0$ & $0$ & $0$ \\
      $C_{24}$ & $0$ & $0$ & $0$ & $0$ & $0$ & $0$ & $0$ \\
      $C_{25}$ & $0$ & $0$ & $0$ & $0$ & $0$ & $0$ & $0$ \\ \hline
      $C_{26}$ & $\not=0$ & $0$ & $0$ & $0$ & $0$ & $0$ & $0$ \\ \hline
      $C_{27}$ & $\not=0$ & $\not=0$ & $\not=0$ & $0$ & $0$ & $0$ & $\not=0$ \\
      $C_{28}$ & $\not=0$ & $\not=0$ & $0$ & $\not=0$ & $0$ & $\not=0$ & $0$ \\
      $C_{29}$ & $\not=0$ & $0$ & $\not=0$ & $\not=0$ & $\not=0$ & $0$ & $0$ \\ \hline
      $C_{30}$ & $\not=0$ & $\not=0$ & $\not=0$ & $0$ & $\not=0$ & $\not=0$ & $\not=0$ \\
      $C_{31}$ & $\not=0$ & $\not=0$ & $0$ & $\not=0$ & $\not=0$ & $\not=0$ & $\not=0$ \\
      $C_{32}$ & $\not=0$ & $0$ & $\not=0$ & $\not=0$ & $\not=0$ & $\not=0$ & $\not=0$ \\ \hline
      $C_{33}$ & $\not=0$ & $\not=0$ & $\not=0$ & $\not=0$ & $\not=0$ & $\not=0$ & $\not=0$ \\ \hline
      $C_{34}$ & $0$ & $0$ & $0$ & $0$ & $0$ & $0$ & $0$ \\
      $C_{35}$ & $0$ & $0$ & $0$ & $0$ & $0$ & $0$ & $0$ \\
      $C_{36}$ & $0$ & $0$ & $0$ & $0$ & $0$ & $0$ & $0$ \\
      $C_{37}$ & $0$ & $0$ & $0$ & $0$ & $0$ & $0$ & $0$ \\ \hline
      $C_{38}$ & $\not=0$ & $0$ & $0$ & $0$ & $0$ & $0$ & $0$ \\
      $C_{39}$ & $\not=0$ & $0$ & $0$ & $0$ & $0$ & $0$ & $0$ \\
      $C_{40}$ & $\not=0$ & $0$ & $0$ & $0$ & $0$ & $0$ & $0$ \\
    \end{tabular}
  \end{ruledtabular}
\end{table}
\begin{table}[ht]
  \caption{\label{table:continuous-invariants-2222-2} The continuous invariants $h_1,\ldots,h_7$ for the sets of equivalence classes based on the discrete invariants for $n=4$, $D=(2,2,2,2)$.}
\footnotesize
  \begin{ruledtabular}
    \begin{tabular}{lccccccc}
      & $h_1$ & $h_2$ & $h_3$ & $h_4$ & $h_5$ & $h_6$ & $h_7$ \\ \hline
      $C_{41}$ & $\not=0$ & $\not=0$ & $\not=0$ & $0$ & $0$ & $0$ & $\not=0$ \\
      $C_{42}$ & $\not=0$ & $\not=0$ & $\not=0$ & $0$ & $0$ & $0$ & $\not=0$ \\
      $C_{43}$ & $\not=0$ & $\not=0$ & $0$ & $\not=0$ & $0$ & $\not=0$ & $0$ \\
      $C_{44}$ & $\not=0$ & $\not=0$ & $0$ & $\not=0$ & $0$ & $\not=0$ & $0$ \\
      $C_{45}$ & $\not=0$ & $0$ & $\not=0$ & $\not=0$ & $\not=0$ & $0$ & $0$ \\
      $C_{46}$ & $\not=0$ & $0$ & $\not=0$ & $\not=0$ & $\not=0$ & $0$ & $0$ \\ \hline
      $C_{47}$ & $0$ & $\not=0$ & $\not=0$ & $0$ & $0$ & $0$ & $0$ \\
      $C'_{47}$ & $\not=0$ & $\not=0$ & $\not=0$ & $0$ & $0$ & $0$ & $\not=0$ \\
      $C_{48}$ & $0$ & $\not=0$ & $0$ & $\not=0$ & $0$ & $0$ & $0$ \\
      $C'_{48}$ & $\not=0$ & $\not=0$ & $0$ & $\not=0$ & $0$ & $\not=0$ & $0$ \\
      $C_{49}$ & $0$ & $0$ & $\not=0$ & $\not=0$ & $0$ & $0$ & $0$ \\
      $C'_{49}$ & $\not=0$ & $0$ & $\not=0$ & $\not=0$ & $\not=0$ & $0$ & $0$ \\ \hline
      $C_{50}$ & $0$ & $0$ & $0$ & $0$ & $0$ & $0$ & $0$ \\
      $C_{51}$ & $0$ & $0$ & $0$ & $0$ & $0$ & $0$ & $0$ \\
      $C_{52}$ & $0$ & $0$ & $0$ & $0$ & $0$ & $0$ & $0$ \\
      $C_{53}$ & $0$ & $0$ & $0$ & $0$ & $0$ & $0$ & $0$ \\ \hline
      $C_{54}$ & $\not=0$ & $0$ & $0$ & $0$ & $0$ & $0$ & $0$ \\
      $C_{55}$ & $\not=0$ & $0$ & $0$ & $0$ & $0$ & $0$ & $0$ \\
      $C_{56}$ & $\not=0$ & $0$ & $0$ & $0$ & $0$ & $0$ & $0$ \\
      $C_{57}$ & $\not=0$ & $0$ & $0$ & $0$ & $0$ & $0$ & $0$ \\
      $C_{58}$ & $\not=0$ & $0$ & $0$ & $0$ & $0$ & $0$ & $0$ \\
      $C_{59}$ & $\not=0$ & $0$ & $0$ & $0$ & $0$ & $0$ & $0$ \\ \hline
      $C_{60}$ & $\not=0$ & $0$ & $0$ & $0$ & $\not=0$ & $\not=0$ & $\not=0$ \\ \hline
      $C_{61}$ & $0$ & $\not=0$ & $\not=0$ & $0$ & $0$ & $0$ & $0$ \\
      $C'_{61}$ & $\not=0$ & $\not=0$ & $\not=0$ & $0$ & $0$ & $0$ & $\not=0$ \\
      $C_{62}$ & $0$ & $\not=0$ & $0$ & $\not=0$ & $0$ & $0$ & $0$ \\
      $C'_{62}$ & $\not=0$ & $\not=0$ & $0$ & $\not=0$ & $0$ & $\not=0$ & $0$ \\
      $C_{63}$ & $0$ & $0$ & $\not=0$ & $\not=0$ & $0$ & $0$ & $0$ \\
      $C'_{63}$ & $\not=0$ & $0$ & $\not=0$ & $\not=0$ & $\not=0$ & $0$ & $0$ \\ \hline
      $C_{64}$ & $0$ & $\not=0$ & $\not=0$ & $0$ & $\not=0$ & $\not=0$ & $\not=0$ \\
      $C'_{64}$ & $\not=0$ & $\not=0$ & $\not=0$ & $0$ & $\not=0$ & $\not=0$ & $0$ \\
      $C''_{64}$ & $\not=0$ & $\not=0$ & $\not=0$ & $0$ & $\not=0$ & $\not=0$ & $\not=0$ \\
      $C_{65}$ & $0$ & $\not=0$ & $0$ & $\not=0$ & $\not=0$ & $\not=0$ & $\not=0$ \\
      $C'_{65}$ & $\not=0$ & $\not=0$ & $0$ & $\not=0$ & $\not=0$ & $0$ & $\not=0$ \\
      $C''_{65}$ & $\not=0$ & $\not=0$ & $0$ & $\not=0$ & $\not=0$ & $\not=0$ & $\not=0$ \\
      $C_{66}$ & $0$ & $0$ & $\not=0$ & $\not=0$ & $\not=0$ & $\not=0$ & $\not=0$ \\
      $C'_{66}$ & $\not=0$ & $0$ & $\not=0$ & $\not=0$ & $0$ & $\not=0$ & $\not=0$ \\
      $C''_{66}$ & $\not=0$ & $0$ & $\not=0$ & $\not=0$ & $\not=0$ & $\not=0$ & $\not=0$ \\
    \end{tabular}
  \end{ruledtabular}
\end{table}
\begin{table}[ht]
  \caption{\label{table:continuous-invariants-2222-3} The continuous invariants $h_1,\ldots,h_7$ for the sets of equivalence classes based on the discrete invariants for $n=4$, $D=(2,2,2,2)$.}
\footnotesize
  \begin{ruledtabular}
    \begin{tabular}{lccccccc}
      & $h_1$ & $h_2$ & $h_3$ & $h_4$ & $h_5$ & $h_6$ & $h_7$ \\ \hline
      $C_{67}$ & $0$ & $\not=0$ & $\not=0$ & $\not=0$ & $\not=0$ & $\not=0$ & $\not=0$ \\
      $C'_{67}$ & $\not=0$ & $\not=0$ & $\not=0$ & $\not=0$ & $\not=0$ & $\not=0$ & $0$ \\
      $C''_{67}$ & $\not=0$ & $\not=0$ & $\not=0$ & $\not=0$ & $\not=0$ & $0$ & $\not=0$ \\
      $C'''_{67}$ & $\not=0$ & $\not=0$ & $\not=0$ & $\not=0$ & $0$ & $\not=0$ & $\not=0$ \\
      $C''''_{67}$ & $\not=0$ & $\not=0$ & $\not=0$ & $\not=0$ & $\not=0$ & $\not=0$ & $\not=0$ \\ \hline
      $C_{68}$ & $\not=0$ & $0$ & $0$ & $0$ & $0$ & $0$ & $0$ \\
      $C_{69}$ & $\not=0$ & $0$ & $0$ & $0$ & $0$ & $0$ & $0$ \\
      $C_{70}$ & $\not=0$ & $0$ & $0$ & $0$ & $0$ & $0$ & $0$ \\
      $C_{71}$ & $\not=0$ & $0$ & $0$ & $0$ & $0$ & $0$ & $0$ \\ \hline
      $C_{72}$ & $0$ & $0$ & $0$ & $0$ & $\not=0$ & $\not=0$ & $\not=0$ \\ \hline
      $C_{73}$ & $0$ & $\not=0$ & $\not=0$ & $0$ & $0$ & $0$ & $0$ \\
      $C_{74}$ & $0$ & $\not=0$ & $\not=0$ & $0$ & $0$ & $0$ & $0$ \\
      $C_{75}$ & $0$ & $\not=0$ & $0$ & $\not=0$ & $0$ & $0$ & $0$ \\
      $C_{76}$ & $0$ & $\not=0$ & $0$ & $\not=0$ & $0$ & $0$ & $0$ \\
      $C_{77}$ & $0$ & $0$ & $\not=0$ & $\not=0$ & $0$ & $0$ & $0$ \\
      $C_{78}$ & $0$ & $0$ & $\not=0$ & $\not=0$ & $0$ & $0$ & $0$ \\ \hline
      $C_{79}$ & $0$ & $\not=0$ & $\not=0$ & $0$ & $\not=0$ & $\not=0$ & $\not=0$ \\
      $C_{80}$ & $0$ & $\not=0$ & $0$ & $\not=0$ & $\not=0$ & $\not=0$ & $\not=0$ \\
      $C_{81}$ & $0$ & $0$ & $\not=0$ & $\not=0$ & $\not=0$ & $\not=0$ & $\not=0$ \\ \hline
      $C_{82}$ & $0$ & $\not=0$ & $\not=0$ & $\not=0$ & $0$ & $0$ & $0$ \\
    \end{tabular}
  \end{ruledtabular}
\end{table}
\begin{table}[ht]
  \caption{\label{table:continuous-invariants-2222-4} The continuous invariants $h_1,\ldots,h_7$ for the sets of equivalence classes based on the discrete invariants for $n=4$, $D=(2,2,2,2)$.}
\footnotesize
  \begin{ruledtabular}
    \begin{tabular}{cccccccl}
      $h_1$ & $h_2$ & $h_3$ & $h_4$ & $h_5$ & $h_6$ & $h_7$ & \\ \hline
      \multirow{2}{*}{$0$} & \multirow{2}{*}{$0$} & \multirow{2}{*}{$0$} & \multirow{2}{*}{$0$} & \multirow{2}{*}{$0$} & \multirow{2}{*}{$0$} & \multirow{2}{*}{$0$} & $C_{0},C_{1},\{C_{2},\ldots,C_{7}\},\{C_{11},\ldots,C_{14}\},\{C_{15},\ldots,C_{18}\}$, \\ & & & & & & & $C_{19},\{C_{20},\ldots,C_{25}\},\{C_{34},\ldots,C_{37}\},\{C_{50},\ldots,C_{53}\}$ \\
      $\not=0$ & $0$ & $0$ & $0$ & $0$ & $0$ & $0$ & $C_{26},\{C_{38},C_{39},C_{40}\},\{C_{54},\ldots,C_{59}\},C_{60},\{C_{68},\ldots,C_{71}\}$ \\
      $0$ & $\not=0$ & $\not=0$ & $0$ & $0$ & $0$ & $0$ & $C_{47},C_{61},C_{73},C_{74}$ \\
      $0$ & $\not=0$ & $0$ & $\not=0$ & $0$ & $0$ & $0$ & $C_{48},C_{62},C_{75},C_{76}$ \\
      $0$ & $0$ & $\not=0$ & $\not=0$ & $0$ & $0$ & $0$ & $C_{49},C_{63},C_{77},C_{78}$ \\
      $0$ & $0$ & $0$ & $0$ & $\not=0$ & $\not=0$ & $\not=0$ & $C_{72}$ \\
      $0$ & $\not=0$ & $\not=0$ & $\not=0$ & $0$ & $0$ & $0$ & $C_{82}$ \\
      $\not=0$ & $\not=0$ & $\not=0$ & $0$ & $0$ & $0$ & $\not=0$ & $C_{8},C_{27},C_{41},C_{42}C'_{47},C'_{61}$ \\
      $\not=0$ & $\not=0$ & $0$ & $0$ & $\not=0$ & $\not=0$ & $0$ & $C_{9},C_{28},C_{43},C_{44},C'_{48},C'_{62}$ \\
      $\not=0$ & $0$ & $\not=0$ & $\not=0$ & $\not=0$ & $0$ & $0$ & $C_{10},C_{29},C_{45},C_{46}C'_{49},C'_{63}$ \\
      $0$ & $\not=0$ & $\not=0$ & $0$ & $\not=0$ & $\not=0$ & $\not=0$ & $C_{64},C_{79}$ \\
      $0$ & $\not=0$ & $0$ & $\not=0$ & $\not=0$ & $\not=0$ & $\not=0$ & $C_{65},C_{80}$ \\
      $0$ & $0$ & $\not=0$ & $\not=0$ & $\not=0$ & $\not=0$ & $\not=0$ & $C_{66},C_{81}$ \\
      $\not=0$ & $\not=0$ & $\not=0$ & $0$ & $\not=0$ & $\not=0$ & $0$ & $C'_{64}$ \\
      $\not=0$ & $\not=0$ & $0$ & $\not=0$ & $\not=0$ & $0$ & $\not=0$ & $C'_{65}$ \\
      $\not=0$ & $0$ & $\not=0$ & $\not=0$ & $0$ & $\not=0$ & $\not=0$ & $C'_{66}$ \\
      $\not=0$ & $\not=0$ & $\not=0$ & $0$ & $\not=0$ & $\not=0$ & $\not=0$ & $C_{30},C''_{64}$ \\
      $\not=0$ & $\not=0$ & $0$ & $\not=0$ & $\not=0$ & $\not=0$ & $\not=0$ & $C_{31},C''_{65}$ \\
      $\not=0$ & $0$ & $\not=0$ & $\not=0$ & $\not=0$ & $\not=0$ & $\not=0$ & $C_{32},C''_{66}$ \\
      $0$ & $\not=0$ & $\not=0$ & $\not=0$ & $\not=0$ & $\not=0$ & $\not=0$ & $C_{67}$ \\
      $\not=0$ & $\not=0$ & $\not=0$ & $\not=0$ & $\not=0$ & $\not=0$ & $0$ & $C'_{67}$ \\
      $\not=0$ & $\not=0$ & $\not=0$ & $\not=0$ & $\not=0$ & $0$ & $\not=0$ & $C''_{67}$ \\
      $\not=0$ & $\not=0$ & $\not=0$ & $\not=0$ & $0$ & $\not=0$ & $\not=0$ & $C'''_{67}$ \\
      $\not=0$ & $\not=0$ & $\not=0$ & $\not=0$ & $\not=0$ & $\not=0$ & $\not=0$ & $C_{33},C''''_{67}$ \\
    \end{tabular}
  \end{ruledtabular}
\end{table}

\section{Conclusions}

Mathematical structure of entangled states gives rise to new entanglement invariants, which lead to a new method of entanglement classification.
The invariants describe algebraic properties of linear maps associated with the states.
For finite-dimensional spaces, each invariant takes a value from a finite set of integers, and the resulting set of entanglement classes is finite.
The relation to the standard continuous invariants is such that different values of the discrete invariants correspond to certain continuous invariants being zero or nonzero.
We believe that our classification is the most refined restricted classification possible.
Although this result is formulated as a theorem in the text, its proof is not a usual mathematical proof, but rather a proof by exhaustion of possibilities.

The new method works for an arbitrary finite number of spaces of finite dimensions.
As its application, we obtained entanglement classifications for a wide selection of individual cases of three subsystems and the case of four qubits.

For three subsystems, in addition to finding classifications for individual values of $D=(d_1,d_2,d)$, it is rather easy to obtain results for infinite sequences of values of $d$.
An interesting general feature of these results (which is easy to prove) is that increasing $d$ beyond $d_1 d_2$ does not introduce any new entanglement classes.
As examples, we have found such classifications for the values $(d_1,d_2)\in\{(2,2),(2,3),(2,4),(2,5),(2,6),(3,3),(3,4)\}$ and arbitrary $d$.
Only one of these sequences, $D=(2,2,d)$, had been conjectured in the literature, for which our method gives the same number of classes as the classifications in \cite{Gelfand}, \cite{Dur}, \cite{Miyake1}, \cite{Miyake2} and the conjectured classification in ~\cite{Miyake1}, \cite{Miyake2}.

Entanglement classes and representative elements could be easily generated for other infinite sequences.
The classification problem for the general case of three subsystems, however, is challenging and currently under study.
Note that the entanglement of a set of three large spin subsystems is in some practical sense complementary to a system of many low spin (e.g., many qubit) subsystems.
Both have potential for the construction of practical devices.

The classification of entanglement of four qubits has been considered by several groups of authors~\cite{Verstraete,Wallach,Lamata,Cao,Li,Akhtarshenas,Chterental,Borsten:2010db}.
All previous works found $9$ or fewer fundamental sets of classes after permutations have been removed.
In our work we found $27$ fundamental sets of classes.
Our refined classification could be useful to experimenters who consider detailed properties of four qubit systems.
For example, Barreiro et al.~\cite{Barreiro} find a rich dynamics when they arrange four $\textrm{Ca}^+$ ions as qubits and study entanglement via decoherence and dissipation.
See also~\cite{experiment} for earlier 4 qubit work.

To deepen our knowledge about other quantum systems, their entanglement should be thoroughly studied as well.
Our method provides a simple, general, practical approach to such studies. 

Our new invariants are topological since they are the dimensions of linear spaces.
Although the invariants are rather simple from the point of view of topology, they may have a different interpretation when viewed from other perspective.
For an example of a possibly related interpretation, see \cite{Kauffman}.
Finally, it is also worth pointing out that while we find pure representative states for each class, it is straightforward to combine them into mixed states via a density matrix approach.

\begin{acknowledgments}
We thank Mike Duff and Dietmar Bisch for useful discussions and encouragements, Robert Feger for help with parallel computing, and an anonymous referee for generous and valuable comments. 
RVB acknowledges support from DOE grant at ASU and from Arizona State Foundation.
The work of TWK was supported by DOE grant number DE-FG05-85ER40226.
\end{acknowledgments}

\end{document}